\documentclass[twocolumn]{article}

\usepackage{natbib}
\usepackage{graphicx}
\usepackage{url}

\begin{document}

\title{Using 3D and 2D analysis for analyzing large-scale asymmetry in galaxy spin directions}

\author{Lior Shamir \\ Kansas State University \\ Manhattan, KS 66506 \\ email: lshamir@mtu.edu}


\date{}

\maketitle

\begin{abstract}

The nature of galaxy spin is still not fully known. Iye et al (2021) applied a 3D analysis to a dataset of bright SDSS galaxies that was used in the past for photometric analysis. They showed that the distribution of spin directions of spiral galaxies is random, providing a dipole axis with low statistical significance of 0.29$\sigma$. However, to show random distribution, two decisions were made, each can lead to random distribution regardless of the real distribution of the spin direction of galaxies. The first decision was to limit the dataset arbitrarily to z$<$0.1, which is a redshift range in which previous literature already showed that random distribution is expected. More importantly, while the 3D analysis requires the redshift of each galaxy, the analysis was done with the photometric redshift. If the asymmetry existed, its signal is expected to be an order of magnitude weaker than the error of the photometric redshift, and therefore the low statistical signal under these conditions is expected. When using the exact same data without limiting to $z_{phot}<0.1$ and without using the photometric redshift, the distribution of the spin directions in that dataset shows a statistical signal of $>2\sigma$. Code and data for reproducing the analysis are publicly available. These results are in agreement with other experiments with SDSS, Pan-STARRS, HST, and the DESI Legacy Survey. The paper also examines other previous studies that showed random distribution in galaxy spin directions. While further research will be required, the current evidence suggest that large-scale asymmetry between the number of clockwise and counterclockwise galaxies cannot be ruled out.


\end{abstract}


\section{Introduction}
\label{introduction}

Spiral galaxies as seen from Earth can seem to an observer to spin clockwise (Z) or counterclockwise (S). Since the spin direction is merely a matter of the perspective of the observer, the null hypothesis would be that in a sufficiently large number of galaxies the number of galaxies spinning clockwise would be equal (within statistical error) to the number of galaxies spinning counterclockwise.

Whether the distribution of spin directions of spiral galaxies is indeed random is a question that has been the focus of several previous studies, some of them  suggested that the distribution is not necessarily random \citep{macgillivray1985anisotropy,longo2011detection,shamir2012handedness,lee2019galaxy,lee2019mysterious,shamir2021large}. 

One of the first studies that proposed a population-based asymmetry between galaxies with opposite spin directions was \citep{macgillivray1985anisotropy}. By using a relatively small dataset of just 418 annotated galaxies, they proposed an asymmetry between the number of galaxies spinning in opposite directions with statistical significance of P=0.08 to occur by chance. More recent studied used the power of digital sky surveys enabled by robotic telescopes, allowing the collect larger datasets.

\cite{lee2019mysterious} identified links between the spin directions of spiral galaxies that are too far from each other to interact gravitationally, and defined these links ``mysterious'', cautiously proposing the possible existence of cosmological-scale links between galaxy spin directions \citep{lee2019mysterious}. These claims are aligned with certain evidence using several different datasets from SDSS \citep{shamir2020patterns}, HST \citep{shamir2020pasa}, Pan-STARRS \citep{shamir2020patterns} and DESI Legacy Survey \citep{shamir2021large}. All of these telescopes show very similar profiles of non-random distribution of galaxy spin directions \citep{shamir2021large,shamir2022large}.  The analysis shows a dipole axis in galaxy spin directions observed in all of these telescopes, and the locations of the axes computed from the different telescopes are well within 1$\sigma$ from each other. A statistically significant correlation was also found between the spin directions of galaxies and cosmic initial conditions, proposing that the galaxy spin direction can be used as a probe to study the early Universe \citep{motloch2021observed}.

A study with a dataset of $\sim6.4\cdot10^4$ Sloan Digital Sky Survey (SDSS) galaxies with spectra showed non-random distribution that can be fitted into cosine dependence with probability of 4.6$\sigma$ \citep{shamir2020patterns}. The profile of distribution was nearly identical to a similar analysis done with Pan-STARRS galaxies, when the redshift distribution of the galaxies was similar \citep{shamir2020patterns}. These results are also in excellent agreement with the distribution of $\sim8\cdot10^5$ galaxies from DESI Legacy Survey \citep{shamir2021large}. Galaxies imaged by Hubble Space Telescope annotated manually also show non-random distribution, and a dipole axis very close to the dipole formed by SDSS galaxies with higher redshifts \citep{shamir2020pasa}.

On the other hand, several studies suggested that the distribution of the spin direction of spiral galaxies was random. One of the early studies that suggested random distribution was \citep{iye1991catalog}, who compiled a catalog of $\sim6.5\cdot10^3$ galaxies from the Southern hemisphere, and found random distribution of their spin directions. Another notable attempt to characterize the distribution of galaxy spin directions was done by using anonymous volunteers who annotated a large number of galaxies manually through a web-based user interface \citep{land2008galaxy}. After correcting the data for the bias driven by the human perception, the study concluded that the spin directions of the galaxies were distributed randomly \citep{land2008galaxy}. A study that applied computer annotation confirmed that the distribution of the galaxies in SDSS was indeed random \citep{hayes2017nature}. These studies will be discussed in detail in Section~\ref{previous_studies} of this paper.

\cite{iye2020spin} proposed that the possible observed asymmetry between galaxies with opposite spin directions is the result of photometric objects that are part of the same galaxies in the dataset \citep{iye2020spin}. Unlike the analysis shown in \citep{shamir2012handedness,shamir2020large,shamir2020patterns,shamir2020pasa,shamir2021particles}, in which the spin directions of the galaxies were fitted into cosine dependence based on their RA and declination, \cite{iye2020spin} performed a 3D analysis that used the RA, declination and redshift \citep{iye2020spin}. While their analysis showed that the distribution of galaxy spin directions was random, a follow-up analysis by the National Astronomical Observatory of Japan using the exact same data showed that when using basic statistics, the distribution of the galaxy spin directions in that dataset is not random \citep{Fukumoto2021}.


This paper analyzes the reasons of the differences between the initial analysis published in \citep{iye2020spin} and the follow-up analysis done using the exact same data but showed non-random distribution. The paper also shows analysis of two different datasets that were not analyzed in that manner in the past, and investigates the reasons for the differences between the conclusions of \cite{iye2020spin} and the results shown in \citep{macgillivray1985anisotropy,longo2011detection,shamir2012handedness,shamir2020patterns,shamir2021particles,shamir2021large,shamir2022new,shamir2022large,shamir2022analysis}. The paper also discusses previous studies that showed random distribution of the galaxy spin directions, and analyzes the experimental design that led to these conclusions.

\section{The datasets}
\label{dataset}

The analysis of the possible asymmetry in the distribution of galaxy spin directions has been studied using relatively large datasets for over a decade. During that time, multiple datasets were prepared and studied. The datasets are different in the type of objects, telescopes, redshift limit, magnitude limit, and galaxy annotation methods. Table~\ref{datasets} shows the datasets used in each study. It also shows the purpose for which each dataset was designed. More details about each dataset can be found in the cited papers. The table only shows datasets used by this author in previous experiments. Data collected or used by others will be described in Sections~\ref{other_datasets} and~\ref{previous_studies}.

\begin{table*}
\scriptsize

\begin{tabular}{|l|c|c|c|c|c|c|}
\hline
Number & Instrument          & Reference         &  Object  & Object & Annotation & Purpose \\
            &                          &                         &  type     & count   & method     &             \\
\hline
1 &  SDSS  & \citep{shamir2012handedness} & Spectroscopic & 126,501 & Automatic & S/Z distribution and dipole axis \\
\hline
2 & SDSS  & \citep{shamir2016asymmetry} & Spectroscopic & 13,440 & Hybrid & S/Z photometric asymmetry  \\
\hline
3 & SDSS  & \citep{shamir2017photometric} & Photometric & 162,514 & Automatic & S/Z photometric asymmetry  \\
\hline
4 & Pan-STARRS  & \citep{shamir2017large} & Photometric & 29,013 & Automatic & S/Z photometric asymmetry  \\
\hline
5 & SDSS  & \citep{shamir2017large} & Spectroscopic & 40,739 & Manual & S/Z photometric asymmetry  \\
\hline
6   & HST & \citep{shamir2020asymmetry}  & Photometric  &  5,122  & Hybrid & S/Z photometric asymmetry \\
\hline
7   & SDSS & \citep{shamir2020patterns}  & Spectroscopic  &  63,693  & Automatic & S/Z distribution and dipole axis \\
\hline
8   & Pan-STARRS & \citep{shamir2020patterns}  & Photometric  &  38,998  & Automatic & S/Z distribution and dipole axis \\
\hline
9   & SDSS   &  \citep{shamir2020large}  & Photometric  &   172,883    & Automatic & S/Z distribution and dipole axis \\
\hline
10   & HST & \citep{shamir2020pasa}  & Photometric  &  8,690    & Manual & S/Z distribution and dipole axis \\
\hline
11 & SDSS & \citep{shamir2020pasa}  &  Spectroscopic &  15,863   & Automatic & S/Z distribution and dipole axis \\
\hline
12 & SDSS & \citep{shamir2021particles} & Photometric  &  77,840  & Automatic & S/Z distribution and dipole axis \\
\hline
13 & DESI Legacy Survey &     \citep{shamir2021large} & Photometric  &  807,898  & Automatic & S/Z distribution and dipole axis \\             
\hline
\end{tabular}
\caption{The different datasets of galaxies separated by their apparent spin direction. The table includes datasets used by this author.}
\label{datasets}
\end{table*}

For their analysis, \cite{iye2020spin} used Dataset 3 in Table~\ref{datasets}, which is a dataset of photometric objects used for the purpose of photometric analysis of objects rotating in opposite directions. In the absence of literature that used that dataset to analyze a dipole axis in the S/Z galaxy distribution, \cite{iye2020spin} compared their results to the results shown in \citep{shamir2020large}, which were based on Dataset 9 in Table~\ref{datasets}. For the comparison, \cite{iye2020spin} argue that the dataset used in \citep{shamir2020large} was the result of combining Dataset 3 in Table~\ref{datasets} with 33,028 galaxies imaged by Pan-STARRS. That statement is made twice in Section 3.1 in the journal version of \citep{iye2020spin}. However, in \citep{shamir2020large} no Pan-STARRS galaxies were used. Moreover, no galaxies from two different telescopes were combined into a single dataset in any other previous study, and consequently no statement about combining data from two different telescopes can be found in any previous paper. It is therefore unclear what led \cite{iye2020spin} to believe that the dataset they used for comparing their results contained Pan-STARRS galaxies. 
The two statements about combining the SDSS and Pan-STARRS galaxies appears only in the journal version of \citep{iye2020spin}, but not in the ArXiv version of that paper.

\section{3D analysis of cosmological-scale anisotropy using the photometric redshift}
\label{photometric_redshift}

The analysis done in \citep{shamir2012handedness,shamir2020large,shamir2020patterns,shamir2020pasa,shamir2021particles,shamir2021large} was two-dimensional, and therefore the position of each galaxy was determined by its RA and declination. The RA and declination are considered accurate. \cite{iye2020spin}, however, applied a 3D analysis. Unlike the 2D analysis done in \citep{shamir2012handedness,shamir2020large,shamir2020patterns,shamir2020pasa,shamir2021particles,shamir2021large,shamir2022new,shamir2022large}, the location of each galaxy in the 3D analysis was determined by its $(l,b,d)$, where d is the distance. As explained in Section 2.1 of \citep{iye2020spin}, the distance $d_i$ of each galaxy {\it i} was determined by $d^i=cz^i/H_{o}$ , where {\it c} is the speed of light, $H_{o}$ is the Hubble constant, and {\it z} is the redshift. Because the vast majority of the galaxies in Dataset 3 in Table~\ref{datasets} do not have spectra, the spectroscopic redshift could not be used for the analysis of that dataset. The redshifts used in \citep{iye2020spin} are the photometric redshifts from the catalog of \citep{paul2018catalog}.

The dataset of \citep{shamir2020patterns} is based on spectroscopic objects identified as galaxies, and therefore all galaxies in that dataset had redshift. \cite{iye2020spin}, however, used a dataset of photometric objects used in \citep{shamir2017photometric}. \cite{iye2020spin} report on two experiments of 3D analysis, one with 162,516 objects and another experiment with 111,867 objects with ``measured redshift'' \citep{iye2020spin}. As discussed in \citep{shamir2017photometric}, less than 11K galaxies in that dataset had spectra. Since less than 11K galaxies in that dataset had spectra, the vast majority of the galaxies in \citep{shamir2017photometric} did not have spectroscopic redshift. As shown in Section 3.1 of \citep{iye2020spin}, the source of the redshift was \citep{paul2018catalog}, which is a catalog of photometric redshift. 


Unlike the redshift measured from the spectra, the photometric redshift is highly inaccurate. The average error of the photometric redshift method used in \citep{paul2018catalog} is $\sim$18.5\%. That error is far greater than the possible signal of less than 1\% shown in \citep{shamir2020large,shamir2020patterns}. When the error of the data is an order of magnitude greater than the expected signal, it is expected that analysis of these data would lead to loss of the signal, and therefore random distribution. 

If the photometric redshift is not systematically biased, it is expected that the error in one direction would balance the error in the opposite direction, and therefore any 3D axis that exists in the data will also exist with the approximate same location when the photometric redshift is used. But because the distances between the galaxies when using the photometric redshift are not the accurate distances, even if the error of the photometric redshift is symmetric, such 3D analysis would lead to weaker signal due to the inaccurate distribution of the galaxies in the 3D space.

In addition to the relatively high error of the photometric redshift, the photometric redshift was determined by using machine learning, and therefore was based on complex data-driven rules that are very difficult to characterize in an intuitive manner. Because it is based on machine learning output of multiple parameters that are not directly related to the spectra, it can also be systematically biased. The general nature of the systematic bias of photometric redshift has been discussed and analyzed in multiple previous studies such as \citep{wittman2009lies,bernstein2010catastrophic,dahlen2013critical,rau2015accurate,tanaka2018photometric}.

The catalog of photometric redshift \citep{paul2018catalog} that was used by \cite{iye2020spin} is also systematically biased. For instance, according to that catalog \citep{paul2018catalog}, the mean photometric redshift of the 289,793 galaxies in the RA range $(180^o-210^o)$ is 0.1606$\pm$0.0002, and the mean photometric redshift of the 289,710 in the RA range $(210^o-240^o)$ is 0.1578$\pm$0.0002. The two-tailed probability to have such difference by chance is $\sim$0.0001. Even after applying a Bonferroni correction to all of the 12 30$^o$ RA slices, the probability is still less than 0.0012. In the RA range $(330^o-360^o)$, the mean photometric redshift of the 323,028 galaxies is 0.1705$\pm$0.0002, which is significantly different (P$<$0.001) from the mean photometric redshift in the two other RA slices.

Since the magnitude of the galaxies in that catalog was limited to $i<18$ in all RA ranges, and no criteria were used to select the galaxies, the difference in the mean distance of the galaxies can be viewed as evidence of cosmological-scale anisotropy. However, these results are more likely driven by the systematic bias of the redshift catalog rather than a true reflection of the large-scale structure. It is possible that such differences in redshift are the results in differences in the cosmic voids in different parts of the sky. However, according to a catalog of cosmic voids in SDSS \citep{mao2017cosmic}, the mean spectroscopic redshifts of the voids in the three RA ranges are 0.4724$\pm$0.0084, 0.4615$\pm$0.0088, and 0.4525$\pm$0.012, for RA ranges $(180^o-210^o)$, $(210^o-240^o)$, and $(330^o-360^o)$, respectively. The differences between the mean spectroscopic redshifts of the cosmic voids are not statistically significant. Also, in the RA range $(300^o-330^o)$ the mean photometric redshift is higher than in the RA range $(210^o-240^o)$, while the mean spectroscopic redshift of the voids is lower. In RA range $(180^o-210^o)$ the mean photometric redshift is lower than in the RA range $(330^o-360^o)$, while the mean spectroscopic redshift of the voids is also lower. That inconsistency suggests that the differences in the mean photometric redshift is not necessarily driven by cosmic voids, but more likely by the systematic bias of the photometric redshift. The photometric redshift is a complex probe for studying subtle cosmological-scale anisotropties. Since it is based on non-intuitive data-driven machine learning rules, it is also difficult to fully profile at a precision level.



The combination of high error and systematic bias makes the 3D analysis with the photometric redshift difficult to predict, and therefore an unreliable probe for studying possible subtle violations of the cosmological isotropy assumption. Statistical observations of cosmological-scale anisotropy that are based on the photometric redshift must therefore be analyzed with extreme caution. As the analysis of \citep{iye2020spin} requires the redshift of each galaxy, the absence or presence of non-random distribution must be analyzed with the spectroscopic redshift rather than the photometric redshift.

\section{Limiting the redshift range}
\label{limiting_redshift}

By applying the 3D analysis using the photometric redshift, \cite{iye2020spin} identified a dipole axis at $(\alpha=26^o,\delta=23^o)$ with statistical significance of $4.00\sigma$. However, the 4.00$\sigma$ was shown when the redshift of the galaxies in the dataset was limited to $z_{phot}<0.1$. As shown in \citep{shamir2020patterns}, there is no statistically significant asymmetry when using galaxies with redshift range of $z<0.1$. Therefore, the dipole axis shown in \citep{iye2020spin} is in strong disagreement with the dipole axis shown in \citep{shamir2020patterns}.   

As shown in previous experiments, if galaxy spin directions are not distributed fully randomly, such distribution might be related to the redshift range \citep{shamir2020patterns,shamir2020pasa}. For instance, it has been shown that when normalizing the redshift distribution, SDSS and Pan-STARRS show a near-identical profile of the S/Z distribution \citep{shamir2020patterns}. Similarly, using SDSS galaxies with relatively high redshift (z$>$0.15) provides a profile of asymmetry that is very similar to the asymmetry of galaxies imaged by HST \citep{shamir2020pasa}. 

As shown in \citep{shamir2020patterns}, if indeed there is an asymmetry between the number of Z and S galaxies, that asymmetry might not be present at lower redshifts. Tables 3, 5, 6, and 7 in \citep{shamir2020patterns} show random distribution at z$<$0.15 \citep{shamir2020patterns}. An analysis of a dipole axis when limiting the redshift to $z<0.15$ showed no statistically significant dipole axis. Therefore, the random distribution at $z_{phot}<0.1$ reported in \citep{iye2020spin} is in full agreement with \citep{shamir2020patterns}. 

When not limiting the redshift, \cite{iye2020spin} report on a dipole axis of 1.29$\sigma$ at $(\alpha=181^o,\delta=31^o)$. The stronger signal when the redshift range is higher is in agreement with \citep{shamir2020patterns}. However, \cite{iye2020spin} used the photometric redshift for a 3D analysis. The reported asymmetry between the number of galaxies with opposite spin directions is $\sim$1\%, and therefore the error of the data was far greater than the expected signal. When analyzing data with error greater than the signal, random distribution is expected. 


\section{Reanalysis of the Iye catalog}
\label{reanalysis}

As discussed in Sections~\ref{photometric_redshift} and~\ref{limiting_redshift}, to show statistical signal of 0.29$\sigma$ \cite{iye2020spin} used the photometric redshift in a 3D analysis, and also limited the dataset to z$<$0.1. Each of these decisions can eliminate the presence of a possible signal. Therefore, evidence of random or non-random distribution of Z and S spiral galaxies can be studied when using the exact same 72,888 galaxies used in \citep{iye2020spin}, but without limiting the dataset by the redshift, and without using 3D analysis that is based on the photometric redshift. The data are available at \url{https://people.cs.ksu.edu/~lshamir/data/assym_72k/}.


A dipole axis in the S/Z distribution means that in one hemisphere there are more Z spiral galaxies than S spiral galaxies, while in the opposite hemisphere there is a higher number of S spiral galaxies than Z galaxies. Table~\ref{hemispheres} shows the number of S-wise and Z-wise galaxies in the hemisphere centered at RA=160$^o$, and in the opposite hemisphere.

\begin{table*}
\scriptsize

\begin{tabular}{|l|c|c|c|c|c|}
\hline
Hemisphere & \# Z-wise          & \# S-wise         &  $\frac{\#Z}{\#S}$  & P                  & P                   \\
   (RA)         &                        &                         &                               & (one-tailed)   & (two-tailed)     \\
\hline
$70^o-250^o$               & 23,037 & 22,442   &   1.0265   &  0.0026   &  0.0052     \\
$>250^o \cup <70^o$   &  13,660 &  13,749  &   0.9935   &  0.29      &  0.58         \\
\hline
\end{tabular}
\caption{The number of Z-wise and S-wise galaxies in the  \cite{iye2020spin} catalog in the RA hemisphere centered at 160$^o$, and in the opposite hemisphere (centered at RA=340$^o$). The dataset is used without redshift limit. The P values are based on binomial distribution such that the probability of a galaxy to have Z-wise or S-wise spin is 0.5.}
\label{hemispheres}
\end{table*}

Statistically significant signal is observed in one hemisphere. The asymmetry in the opposite, less populated, hemisphere is not statistically significant. But because it has more S-wise galaxies than Z-wise galaxies, it is also not in conflict with the distribution in the hemisphere centered at (RA=160$^o$) for forming a dipole axis in the dataset. That simple analysis shows certain evidence that the distribution in the specific dataset used in \citep{iye2020spin} might not be random. Due to the deterministic nature of the algorithm, repeating the same experiment after mirroring the galaxy images using the ``flip'' command of {\it ImageMagick} led to identically inverse results. For instance, after mirroring the galaxy images the hemisphere $70^o-250^o$ had 22,442 galaxies spinning clockwise, and 23,037 galaxies rotating counterclockwise.

One of the analyses in \citep{iye2020spin} is a simple difference between the number of galaxies that spin clockwise and the number of galaxies that spin counterclockwise. That analysis, however, is done for the entire sky, without separating the sky into two opposite hemispheres. When just counting galaxies in the entire sky, the asymmetry in one hemisphere offsets the asymmetry in the opposite hemisphere. The asymmetry of 1.8$\sigma$ observed in \citep{iye2020spin} can be attributed to the higher number of galaxies in one hemisphere, leading to asymmetry in the total number of galaxies. As shown in Table~\ref{hemispheres}, when separating the sky into two opposite hemispheres the signal becomes statistically significant. As shown in \citep{shamir2021large}, when the number of galaxies is high, the asymmetry in one hemisphere is nearly exactly inverse to the asymmetry in the opposite hemisphere.


As was done in \citep{shamir2012handedness,shamir2020pasa,shamir2020patterns}, from each possible $(\alpha,\delta)$ combination in the sky, the angular distance between that point to all galaxies in the dataset was computed. That was done by using standard angular distance between two points on a sphere. The angular distance $\phi $ between $(\alpha,\delta)$ and galaxy $\Psi$ is determined simply by $\phi=acos(   sin(\delta) \cdot sin(\Psi_{\delta})  + cos(\delta) \cdot cos(\Psi_{\delta}) \cdot cos(\alpha-\Psi_{\alpha}) )$. That analysis does not require the redshift, and therefore can be done with galaxies that do not have spectra.

Then, $\chi^2$ statistics was used to fit the spin direction distribution to cosine dependence. That was done by fitting  $d\cdot|\cos(\phi)|$ to $\cos(\phi)$, where $\phi$ is the angular distance between the galaxy and $(\alpha,\delta)$, and {\it d} is a number within the set $\{-1,1\}$, such that d is 1 if the galaxy spins clockwise, and -1 if the galaxy spins counterclockwise. The $\chi^2$ was compared to the average of the $\chi^2$ when computed in $10^3$ runs such that the $d$ of each galaxy was assigned a random number within \{-1,1\}. The standard deviation of the $\chi^2$ of the $10^3$ runs was also computed. Then, the $\sigma$ difference between the $\chi^2$ computed with the actual spin directions and the $\chi^2$ computed with the random spin directions provided the probability to have a dipole axis in that $(\alpha,\delta)$ combination by chance.

\begin{figure}[h]
\includegraphics[scale=0.14]{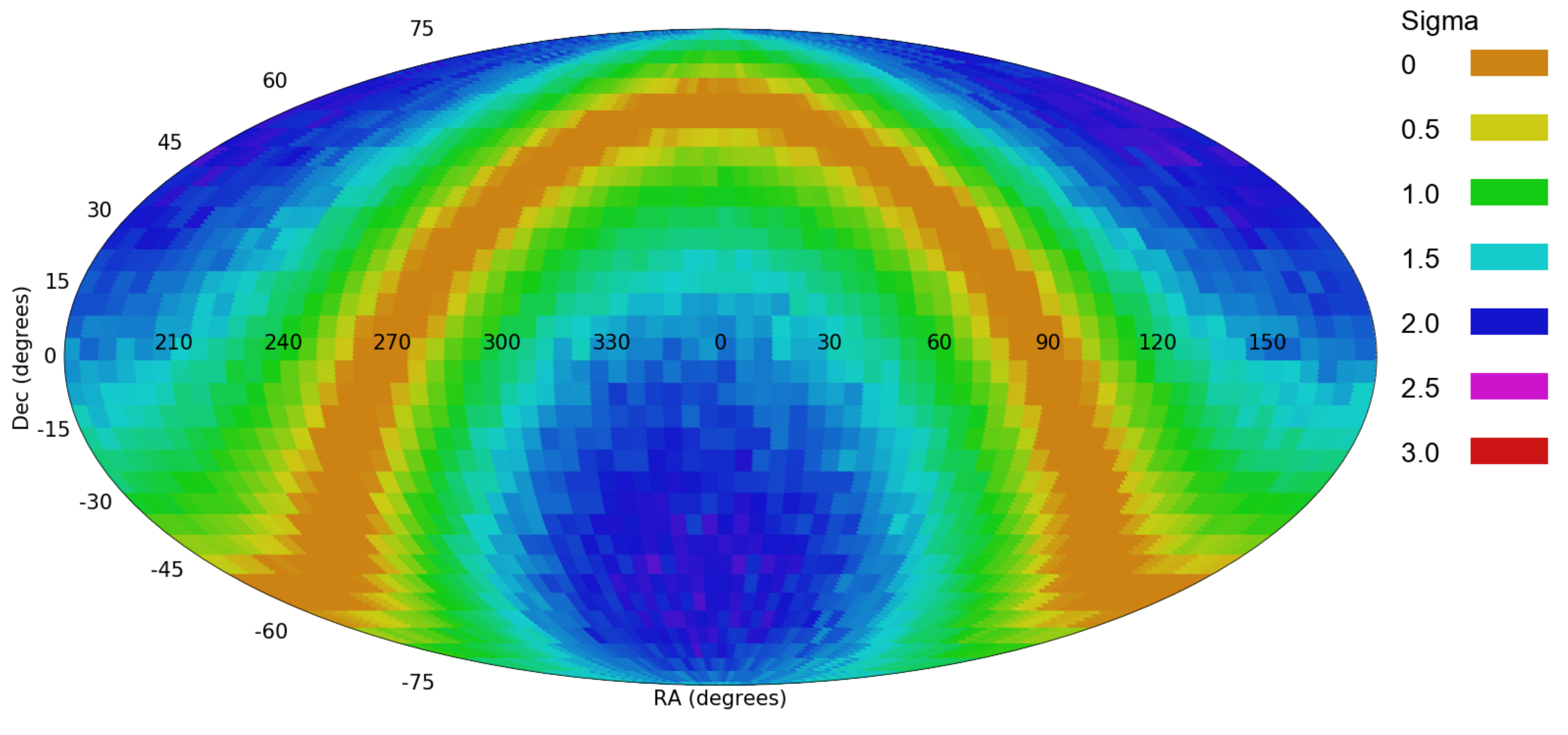}
\caption{The $\chi^2$ probability of a dipole axis in the spin directions of the galaxies from different $(\alpha,\delta)$ combinations.}
\label{dipole1}
\end{figure}

\begin{figure}[h]
\includegraphics[scale=0.15]{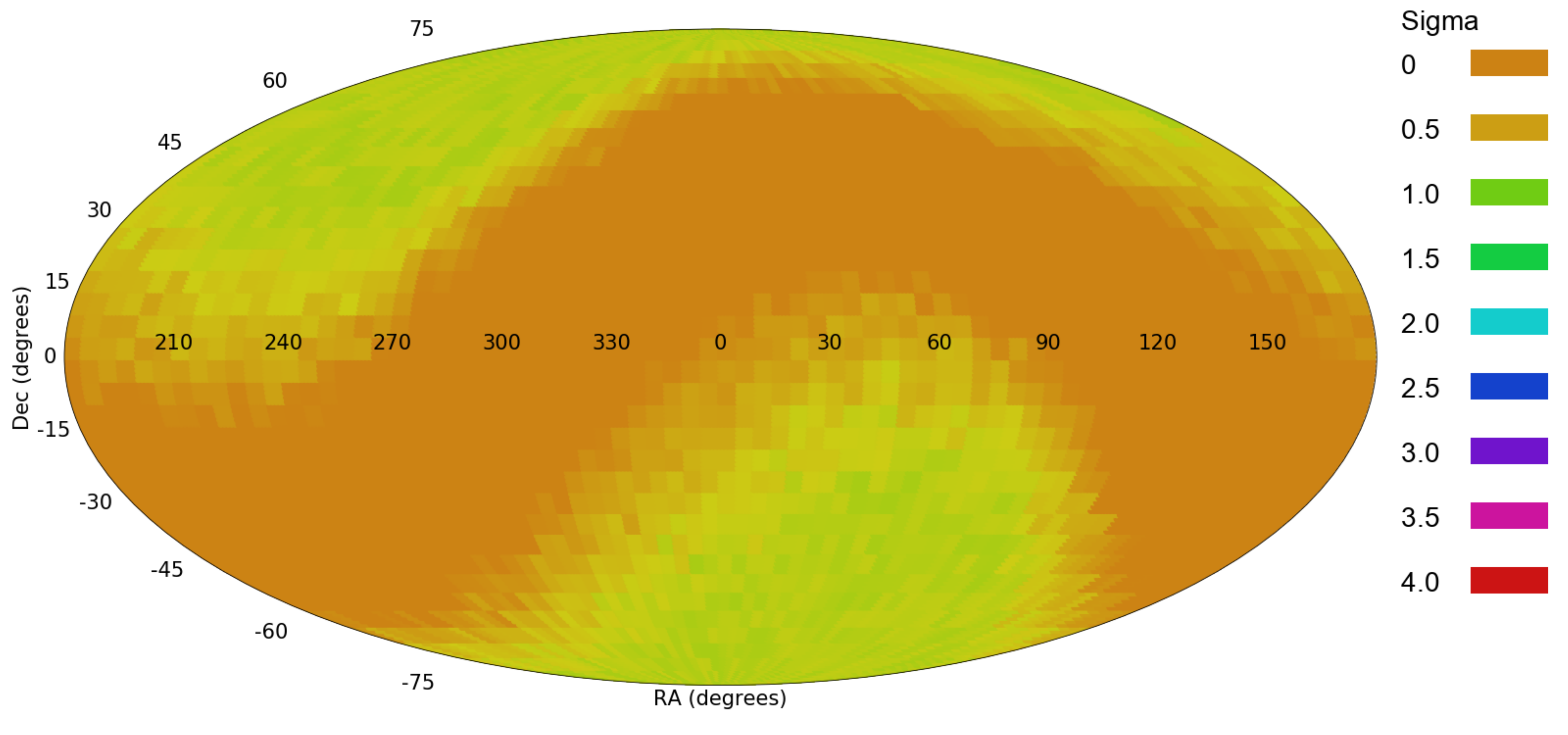}
\caption{The $\chi^2$ probability of the dipole axis when the galaxies are assigned with random spin directions.}
\label{dipole_random}
\end{figure}

Figure~\ref{dipole1} shows the statistical significance of the dipole axis from each possible pair of ($\alpha$,$\delta$) in increments of five degrees. The most likely location of the dipole axis identified at $(\alpha=165^o,\delta=40^o)$, and the probability of the axis is $\sim$2.1$\sigma$. The 1$\sigma$ error range of the position of the axis is $(84^o, 245^o)$ for the right ascension, and $(-41^o, 90^o)$ for the declination. Figure~\ref{dipole_random} shows the likelihood of the dipole axis when the galaxies are assigned with random spin directions, showing much lower probability of $<1\sigma$.

\begin{figure}[h]
\includegraphics[scale=0.15]{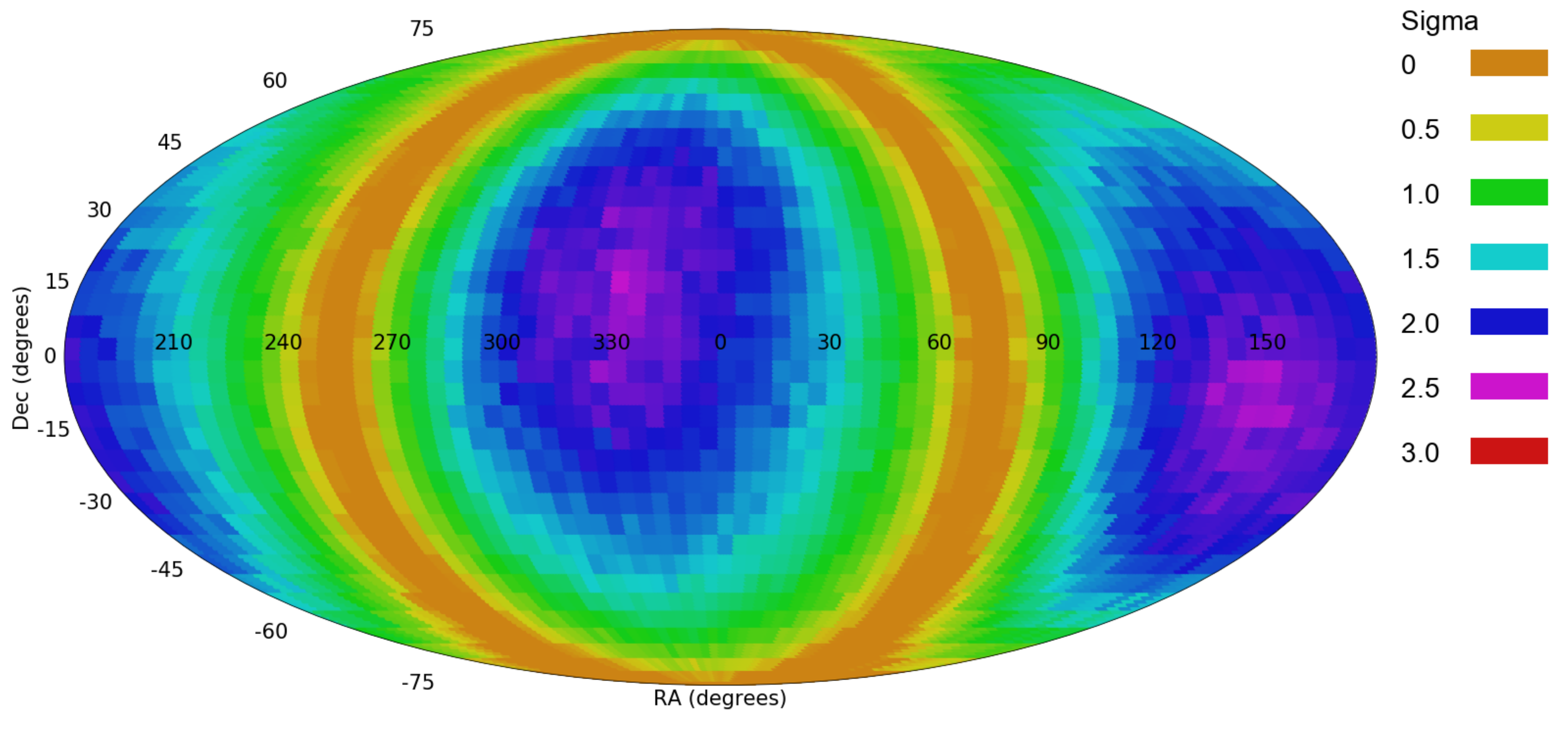}
\caption{The $\chi^2$ probability of a dipole axis in the spin directions of the galaxies from different $(\alpha,\delta)$ combinations when the galaxy population is limited to $z_{phot}>0.1$.}
\label{dipole1_high}
\end{figure}

\begin{figure}[h]
\includegraphics[scale=0.15]{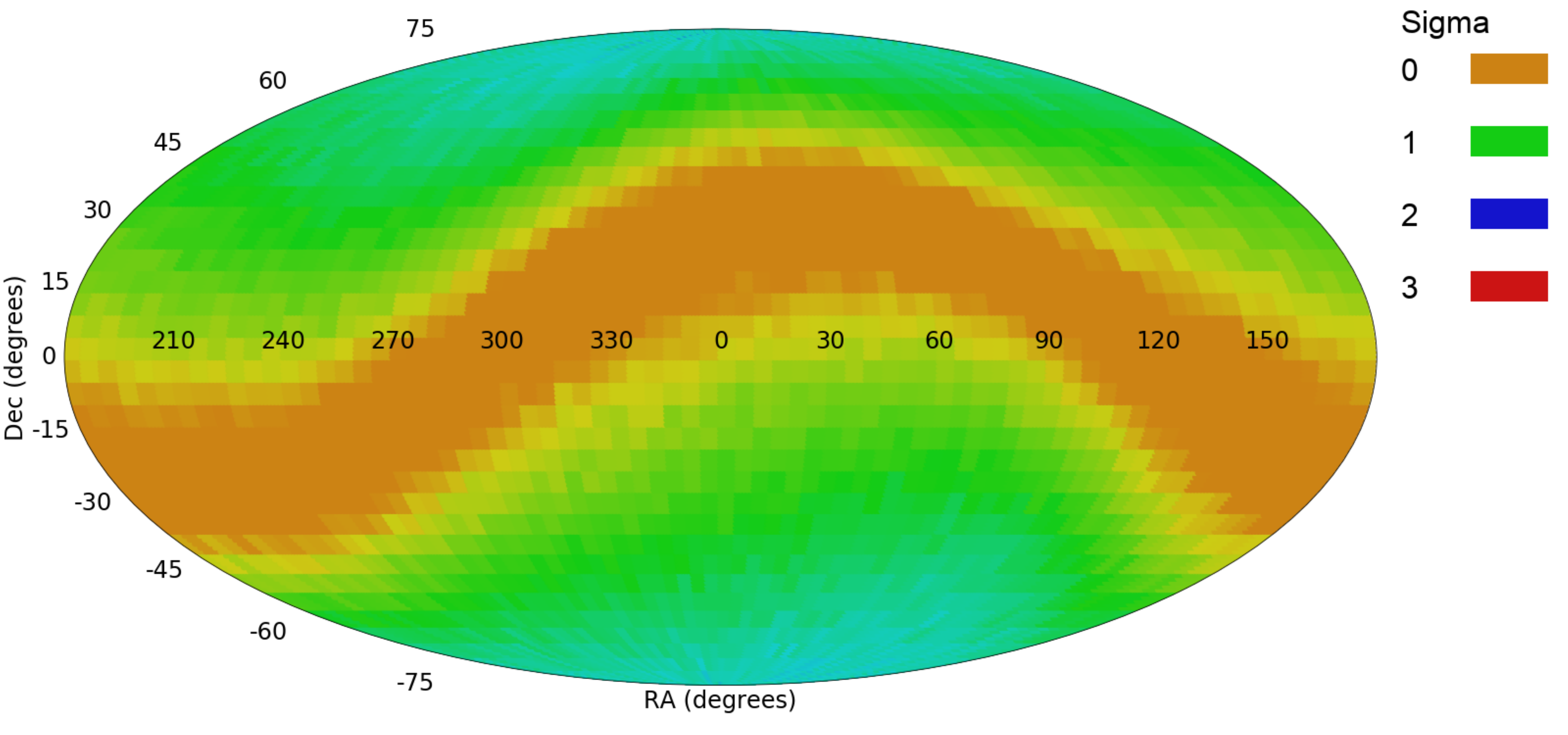}
\caption{The $\chi^2$ probability of a dipole axis in different $(\alpha,\delta)$ combinations when the galaxy population is $z_{phot}<0.1$.}
\label{dipole1_low}
\end{figure}

Figure~\ref{dipole1_high} shows the results of the same analysis when using just the 24,799 galaxies with $z_{phot}>0.1$. When using the galaxies with the higher redshift, the most likely dipole axis is identified in $(\alpha=145^0,\delta=-10^o)$, but the statistical signal increases to $\sim$2.42$\sigma$.  Figure~\ref{dipole1_low} shows the same analysis when the galaxies are limited to $z_{phot}<0.1$, with signal that is not statistically significant. The increased signal when the redshift gets higher is in agreement with the results reported in \citep{shamir2020patterns,shamir2022new,shamir2022large}. As discussed in \citep{shamir2022new,shamir2022large}, one explanation to the change in location and strength of the signal is that such axis, if exists, does not necessarily go directly through Earth.

\begin{figure}[h]
\includegraphics[scale=0.15]{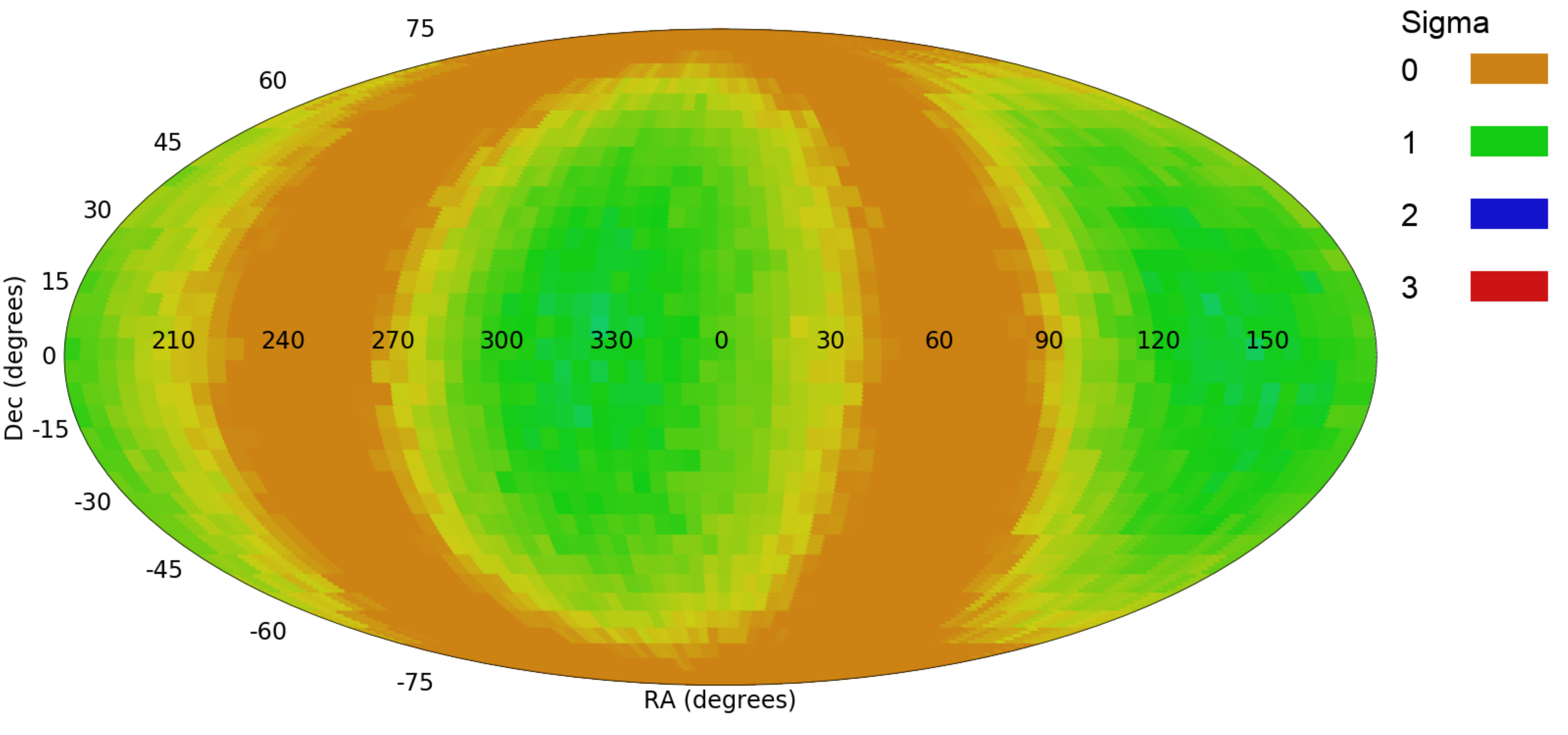}
\caption{The $\chi^2$ probability of a dipole axis in different $(\alpha,\delta)$ combinations when an error of 18.5\% was added to the location of each galaxy. To ensure that the error is symmetric, the direction of the error was assigned randomly.}
\label{with_18prcnt_error}
\end{figure}

Another experiment aimed at testing the impact of the inaccuracy of the locations of the galaxies on the results. For that purpose, an error of 18.5\% was added to the RA and Dec of each galaxy in the dataset. To ensure that the error is added in a symmetric manner, the direction of the error was random. Figure~\ref{with_18prcnt_error} shows the analysis of the dipole axis after adding the error to the locations of the galaxies. While the change of the location of the most likely axis at $(\alpha=155^o,\delta=15^o)$ is minor, the statistical strength of the axis is dropped to 1.41$\sigma$. That shows that an error added randomly to the locations of the galaxies might not change the location of the dipole axis, but the error reduces the signal. 

\section{Error in the galaxy annotation}
\label{error}

The statistical signal observed in the analysis of Section~\ref{reanalysis} can also be the result of the annotation of the galaxies, as even a subtle bias in the annotation process leads to very strong statistical signal \citep{shamir2021particles}. The algorithm used for annotating the galaxies \citep{shamir2011ganalyzer} is fully symmetric. It works by using clear defined rules that are completely symmetric. It is a model-driven algorithm that is not based on machine learning or deep learning algorithms that cannot be fully symmetric due to their sensitivity to the data they are trained by, the initial weights, and even the order of the training samples. Deep learning algorithms might also be driven by biases that are very difficult to characterize \citep{dhar2021evaluation}, specifically in the case of galaxy images \citep{dhar2022systematic}. Because the algorithm is symmetric, inaccurate annotations are expected to impact galaxies that spin clockwise in the same way it impacts galaxies that spin counterclockwise.

\begin{figure}[h]
\includegraphics[scale=0.60]{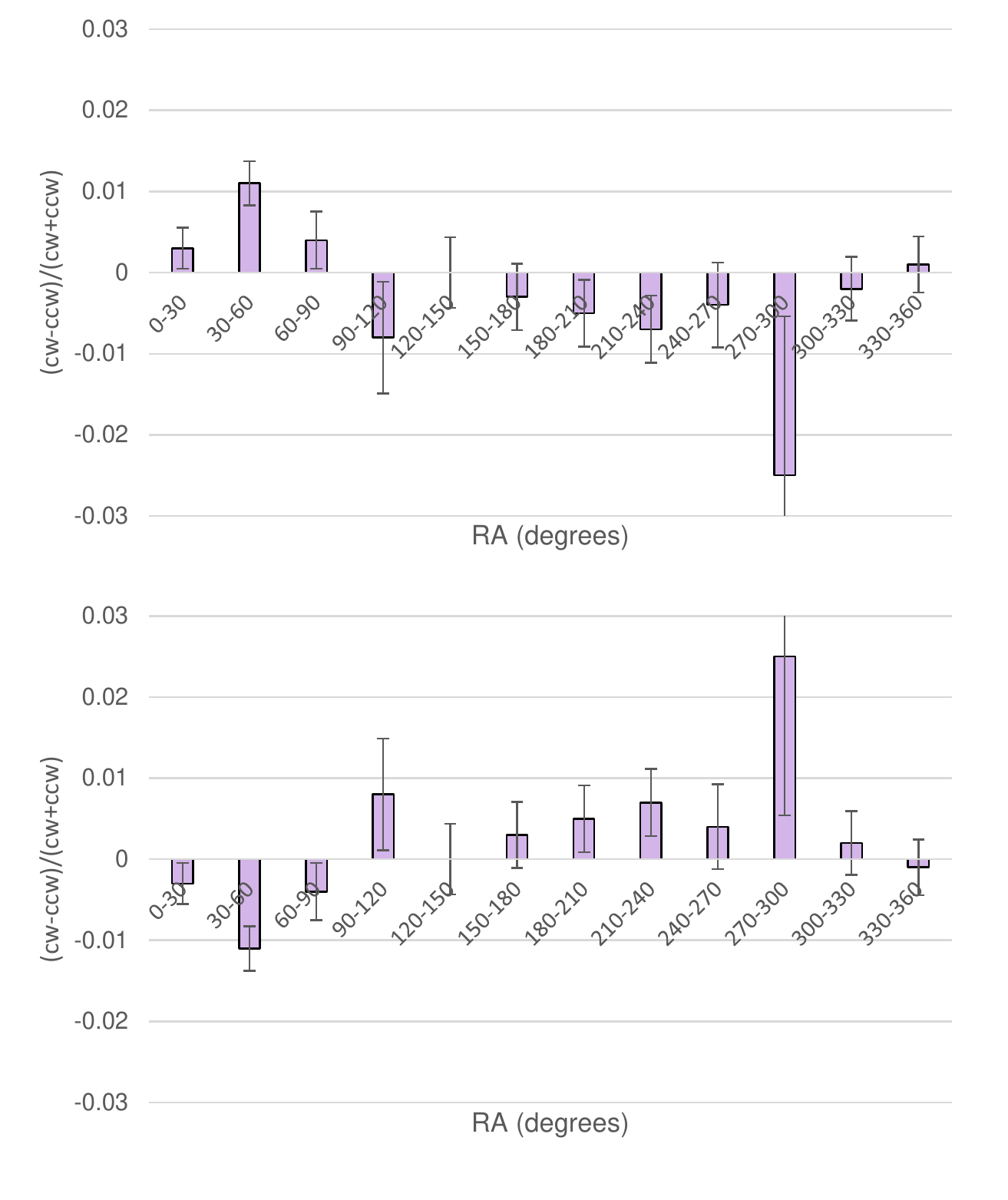}
\caption{Asymmetry in the distribution of the galaxy spin directions in different RA ranges using $\sim8.08\cdot10^5$ galaxies imaged be DECam. The top graph shows the asymmetry observed when analyzing the original images, and the bottom graph shows the asymmetry observed when analyzing the mirrored images.}
\label{by_ra_mirror}
\end{figure}

Several experiments were performed in previous work by repeating the analyses after mirroring the galaxy images \citep{shamir2012handedness,shamir2020large,shamir2020patterns,shamir2020pasa,shamir2021large,shamir2022new,shamir2022large}. In all cases, the results were identically inverse compared to the results with the original images. For instance, Section 2.2 in \citep{shamir2022large} discusses an experiment in which all galaxy images were mirrored, and the results were exactly inverse compared to the results with the original images. The same was done with a far larger dataset as discussed in Section 3 in \citep{shamir2021large}. Figure~\ref{by_ra_mirror} shows the asymmetry in the distribution of 807,898 galaxies imaged by DECam. The experiment is described in \citep{shamir2021large}. The top graph shows the asymmetry observed when analyzing the original images, while the bottom graph shows the asymmetry observed after mirroring the images using the {\it flip} command of the {\it ImageMagick} software tool, and using the lossless TIF image file format. The asymmetry observed with the mirrored galaxies is inverse to the asymmetry observed with the original images.

Another empirical evidence that shows that the observed asymmetry is not the result of bias in the annotations is the inverse asymmetry observed in opposite hemispheres. The inverse asymmetry in opposite hemispheres has been described in previous reports \citep{shamir2020patterns,shamir2021large,shamir2022new,shamir2022large,shamir2022analysis}, and can also be observed in Figure~\ref{by_ra_mirror}. If the algorithm was biased systematically, the same asymmetry should have been observed in all parts of the sky, and it was not expected to inverse in opposite hemispheres as shown in Table~\ref{hemispheres}. That evidence is added to the multiple experiments of mirroring the galaxy images, leading to exactly inverse results \citep{shamir2020patterns,shamir2021large,shamir2022new,shamir2022large}.

If the galaxy annotation algorithm had a certain error in the annotation of the galaxies, the asymmetry {\it A} can be defined by Equation~\ref{asymmetry}.
\begin{equation}
A=\frac{(N_{cw}+E_{cw})-(N_{ccw}+E_{ccw})}{N_{cw}+E_{cw}+N_{ccw}+E_{ccw}},
\label{asymmetry}
\end{equation}
where $E_{cw}$ is the number of Z galaxies incorrectly annotated as S galaxies, and $E_{ccw}$ is the number of S galaxies incorrectly annotated as Z galaxies. Because the algorithm is symmetric, the number of S galaxies incorrectly annotated as Z is expected to be roughly the same as the number of Z galaxies missclassified as S galaxies, and therefore $E_{cw} \simeq E_{ccw}$ \citep{shamir2021particles}. Therefore, the asymmetry {\it A} can be defined by Equation~\ref{asymmetry2}.

\begin{equation}
A=\frac{N_{cw}-N_{ccw}}{N_{cw}+E_{cw}+N_{ccw}+E_{ccw}}
\label{asymmetry2}
\end{equation}

Since $E_{cw}$ and $E_{ccw}$ cannot be negative, a higher rate of incorrectly annotated galaxies is expected to make {\it A} lower. Therefore, incorrect annotation of galaxies is not expected to lead to asymmetry, and can only make the asymmetry lower rather than higher.

An experiment \citep{shamir2021particles} of intentionally annotating some of the galaxies incorrectly showed that even when an error is added intentionally, the results do not change significantly even when as many as 25\% of the galaxies are assigned with incorrect spin directions, as long as the error is added to both Z and S galaxies \citep{shamir2021particles}. But if the error is added in an asymmetric manner, even a small asymmetry of 2\% leads to a very strong asymmetry, and a dipole axis that peaks exactly at the celestial pole \citep{shamir2021particles}.

As described in \citep{shamir2021large}, not all galaxies are spiral, and not all spiral galaxies have a spin direction that can be identified visually. An example of spiral galaxies that their spin direction cannot be determined are spiral edge-on galaxies. Therefore, the spin directions of many of the galaxies cannot be identified, and these galaxies cannot be used in the analysis. A potential source of bias could be when the number of galaxies that their spin direction is not identified is not distributed evenly between galaxies that spin clockwise and galaxies that spin counterclockwise. That is, the number of galaxies that in reality spin clockwise and are also identified by the algorithm as galaxies that spin clockwise can be higher than the number of galaxies that in reality spin counterclockwise and also identified as galaxies that spin clockwise. In that case, the dataset of annotated galaxies can be completely clean, but still exhibit a bias due to the selection of a higher number of galaxies that spin clockwise.

\begin{figure}[h]
\includegraphics[scale=1.10]{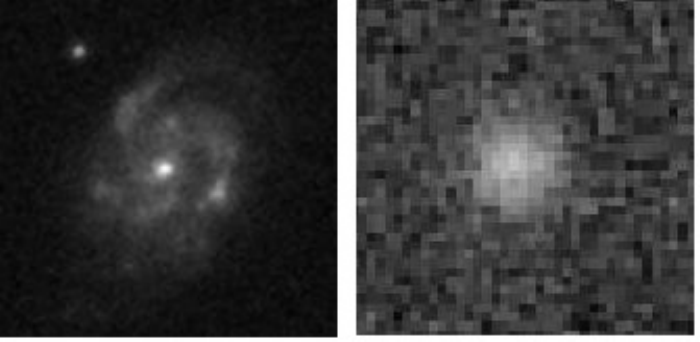}
\caption{A galaxy imaged by HST (left) and the same galaxy imaged by SDSS. The HST image shows a spiral galaxy that spins counterclockwise, while the SDSS image does not allow to identify any kind of spin patterns. The galaxy is at $(\alpha=150.329^o, \delta=1.603^o)$.}
\label{sdss_vs_hst}
\end{figure}

If a galaxy has obvious visible spin patterns, it can be determined that the galaxy spins. However, if a galaxy does not have clear identifiable spin pattern, that does not necessarily mean that the galaxy does not have a clear spin direction. For instance, Figure~\ref{sdss_vs_hst} shows an example of a galaxy at $(\alpha=150.329^o, \delta=1.603^o)$ imaged by HST and by SDSS. The SDSS image of the galaxy does not have an identifiable spin direction, and the galaxy in that image does not necessarily seem spiral. The more powerful HST shows that the galaxy is spiral, with very clear counterclockwise patterns. That example shows that it is practically not possible to separate between galaxies that have spin patterns and galaxies that do not seem to spin. Therefore, such analysis has to rely on the characterization of the data. The algorithm used for a study of this kind must be fully symmetric, as described above.

 Assuming that an annotation algorithm selects a higher number of galaxies from a certain spin direction, it is possible to predict the expected distribution of the spin direction. Given that some galaxies spin in a certain direction but are determined by the algorithm to have an unidentifiable spin direction, the observed asymmetry A can be defined by Equation~\ref{asymmetry3}
\begin{equation}
A= \frac{ R_{cw} \cdot u_{cw} - R_{ccw} \cdot u_{ccw}  } { R_{cw} \cdot u_{cw} + R_{ccw} \cdot u_{ccw}  }  ,
\label{asymmetry3}
\end{equation}
where $R_{cw}$ and $R_{ccw}$ are the numbers of galaxies in the dataset that indeed spin clockwise and counterclockwise, respectively. $u_{cw}$ is the fraction of galaxies in the dataset that spin clockwise, and their spin direction is also identified correctly by the algorithm and therefore these galaxies were used in the analysis. Similarly, $u_{ccw}$ is the fraction of galaxies in the dataset that spin counterclockwise, and their spin directions were identified by the annotation algorithm. 

If the real distribution of the spin directions in the dataset is fully symmetric, we can assume that $R_{cw} \simeq R_{ccw}$. In that case, the observed asymmetry $A$  is defined by Equation~\ref{asymmetry4}.

\begin{equation}
A= \frac{ (u_{cw} - u_{ccw})  } { (u_{cw} + u_{ccw})  }  ,
\label{asymmetry4}
\end{equation}

Because the algorithm does not change during the analysis, the asymmetry $A$ is expected to be a constant. As shown in Figure~\ref{by_ra_mirror} and in several previous experiments \citep{shamir2020patterns,shamir2021large,shamir2022new,shamir2022large}, the asymmetry changes consistently in different parts of the sky. Moreover, the sign of the asymmetry flips in opposite hemispheres. For instance, Table~\ref{hemispheres2} shows the number of clockwise and counterclockwise galaxies imaged by DECam in opposite hemispheres, as described in \citep{shamir2021large}. The table shows statistically significant inverse asymmetry observed in opposite hemispheres. Because the algorithm does not change during the analysis, a selection bias would have been expected to show a consistent asymmetry $A$ in all parts of the sky. It is difficult to think of an algorithm that does not change during the analysis, but selects a higher number of clockwise galaxies in one hemisphere, and a higher number of counterclockwise galaxies in the opposite hemisphere.

\begin{table*}
\caption{The number of DECam galaxies spinning clockwise and counterclockwise in opposite hemispheres. The P values are the binomial distribution probabilities to have such asymmetry or stronger when the probability of a galaxy to spin in each direction is mere chance 0.5 \citep{shamir2021large}. The vast majority of the galaxies are from the Southern hemisphere.}
\label{hemispheres2}
\centering
\begin{tabular}{lcccc}
\hline
\hline
Hemisphere (degrees)      & \# cw galaxies & \# ccw galaxies  & $\frac{cw-ccw}{cw+ccw}$  & P \\
\hline
$(0^o-150^o \cup 330^o-360^o)$ &   264,707    &  262,559  &   0.004      &  0.0015  \\   
$(150^o-330^o)$                          &   139,719     & 140,913  &   -0.004    &  0.0121   \\
\hline
\hline
\end{tabular}
\end{table*}

\subsection{Other biases in the selection of the galaxies}
\label{selection_bias}

A possible source of error or bias can be the selection of the galaxies. Using some selection criteria or previous catalogs of galaxy morphology can lead to bias carried over from the catalog, and affecting the rest of the analysis. For instance, a certain galaxy morphology catalog can be biased due to human or machine learning bias. Both of these possible biases are very difficult to detect and profile. The human perception is complex to understand or reproduce, and machine is based on complex non-intuitive data-driven rules are difficult to control and verify formally that no bias exists. For instance, it has been shown that non-even distribution of the locations of the training samples in the sky can lead to mild but consistent bias in deep neural networks \citep{dhar2022systematic}. 

To annotate the galaxies by their spin directions, one might apply a first step of using machine learning or crowdsourcing to identify spiral galaxies, and then apply an algorithm to the galaxies classified as spiral to identify their spin direction \citep{hayes2017nature}. While this approach does not necessarily lead to a biased dataset, it also introduces a risk that bias in the machine learning algorithm that is used to select spiral galaxies might be carried over to the rest of the analysis. That is, if the selection of the spiral galaxies is biased in some way due to machine learning or human bias, that bias can affect the entire analysis. Such bias might be unintuitive and difficult to identify. 

Because machine learning algorithms use all possible information that can differentiate between the different classes in the training set, these algorithms can be biased by aspects that are not intuitive or obvious to the user \citep{dhar2021evaluation}. For instance, a machine learning algorithm trained to classify between elliptical and spiral galaxies has been shown to also ``learn'' the background sky \citep{dhar2022systematic}. As a result, the distribution of elliptical and spiral galaxies in one part of the sky was different from the distribution in another part of the sky, and the different distributions corresponded to the part of the sky from which the training galaxies were taken. That is, the machine learning algorithm showed a certain distribution of elliptical and spiral galaxies. Then, the same set of galaxies was annotated by the machine learning algorithm, and provided a significantly different distribution of spiral and elliptical galaxies. That was because the machine learning algorithm was trained by different sets of spiral and elliptical galaxies, taken from different parts of the sky. The experiment showed that by selecting training samples from different parts of the sky to train the machine learning algorithm, the results were significantly different, showing two different distributions when annotating the exact same set of galaxies.

To avoid such possible biases, the galaxies should be selected without prior analysis, and in particular avoid complex and unexplainable analyses such as deep neural networks or other forms of machine learning. In \citep{shamir2020patterns}, which is Dataset 7 in Table~\ref{datasets}, all galaxies with spectra in SDSS DR 14 were used. The only selection criterion of the galaxies was the Petrosian radius. All SDSS galaxies with spectra and Petrosian radius greater than 5.5'' were selected and analyzed. Assuming that galaxies that spin clockwise are, on average, as large as galaxies that spin counterclockwise, this simple criterion is not expected to lead to bias in the data. Then, the Ganalyzer algorithm was applied on all galaxies without any prior selection. Galaxies that did not have identified spin direction, such as elliptical galaxies, were rejected from the analysis when Ganalyzer was not able to determine their spin direction. As shown in \citep{shamir2011ganalyzer}, Ganalyzer can identify between spiral and elliptical galaxies, and does that in a fully symmetric manner, without using any form of machine learning or pattern recognition.


In the dataset of DESI Legacy Survey objects used in \citep{shamir2021large}, which is Dataset 13 in Table~\ref{datasets}, the galaxies did not have spectra. In that case, the only selection criteria was to select just extended objects that have r magnitude brighter than 19.5. That selection criteria was necessary due to the extreme size of the data. Assuming that galaxies that spin clockwise are, on average, as bright as galaxies that spin counterclockwise, such simple selection is not expected to lead to preference of galaxies that spin in a certain direction. That selection does not involve machine learning or human selection that can be biased in a manner that is difficult to notice and profile. A similar selection criteria was used in Dataset 8 of Pan-STARRS galaxies. The galaxies were selected from the entire Pan-STARRS dataset, such that the selection criteria was extended objects with r magnitude of less than 19 \citep{timmis2017catalog}. No other selections such as morphological catalogs were used.

As also mentioned in Section~\ref{dataset}, the dataset used by \citep{iye2020spin} was not designed or used to analyze the asymmetry in the population of galaxies spinning in opposite directions. The paper from which the data were taken \citep{shamir2017photometric} makes no attempt to identify any kind of axis in galaxies with opposite spin directions, and no attempt to show a dipole axis in that dataset was made in any other paper. The dataset used in \citep{iye2020spin} is a dataset of galaxies that were selected after a first step of applying machine learning to separate spiral galaxies from elliptical galaxies \citep{shamir2017photometric}. That step makes the dataset different from the datasets mentioned above. 

The machine learning algorithm used to classify the galaxies \citep{orlov2008wnd} was initially developed for classifying cells in microscopy images. Therefore, the features it uses are rotationally invariant. Unlike deep neural networks, where full rotational invariance is more difficult to control, substantial work has been done in the past to develop specific mathematical descriptors that can reflect the visual content and are not affected when the image is rotated or mirrored. Therefore, the machine learning algorithm is not expected to be sensitive to the spin direction of the galaxy. However, machine learning works by complex rules that are not intuitive to understand, and can therefore lead to biases that are not expected. It is therefore preferred not to use machine learning in any step of such analysis. As mentioned above, the dataset used by \cite{iye2020spin} to show a dipole axis in the distribution of galaxy spin directions was not used for that purpose in the paper from which it was taken \citep{shamir2017photometric}, or in any other paper. Experiments that show asymmetry in the distribution of galaxy spin directions \citep{shamir2020patterns,shamir2020pasa, shamir2021large,shamir2022new,shamir2022large} did not use any kind of machine learning, and the selection of the galaxies was done by using very simple rules. In all of these cases there was no first step of selecting spiral galaxies, which is a step that in itself might lead to bias if done by human annotation or machine learning.

\section{Comparing to other datasets}
\label{other_datasets}

The results shown in Section~\ref{reanalysis} can be compared to results published in previous studies. One of the first attempts to use a large number of galaxies to study the distribution of galaxy spin directions was Galaxy Zoo \citep{land2008galaxy}, which analyzed galaxies manually by several hundred thousands non-expert volunteers. That effort showed non-significant distribution that peaked at $(\alpha=161^o,\delta=11^o)$, as also mentioned in \citep{longo2011detection}. That location is close to the location of the most probable axis shown in Section~\ref{reanalysis}. The axis reported with Galaxy Zoo data is not statistically significant. The absence of statistical signal can also be attributed to the strong human bias \citep{land2008galaxy}, leading to substantial corrections during which most of the data could not be used. The average redshift of the galaxies in that dataset was $\sim$0.07.

The main downside of the Galaxy Zoo annotations was that the galaxies were not mirrored randomly, and therefore the perceptional bias of the volunteers did not offset. Section~\ref{previous_studies} provides an additional discussion regarding the Galaxy Zoo experiment. Another attempt to study the possible asymmetry was done by \cite{longo2011detection}, who used 15,158 manually annotated galaxies. These galaxies were mirrored randomly to offset for the possible human bias. The study resulted in a dipole axis in the galaxy spin directions with the most probable axis at $(\alpha=217^o,\delta=32^o)$, with very strong signal of $>5\sigma$. That location falls within the 1$\sigma$ error of the most likely axis shown in Section \ref{reanalysis}, which is $(84^o, 245^o)$ for the RA. The declination of the two axes is nearly identical. The downside of that study was that the galaxies were annotated manually by five undergraduate students, and therefore the annotations are subjected to possible biases related to human labor.

\cite{longo2011detection} does not specify the exact average redshift of the galaxies, but specifies that the redshift of the galaxies was limited to 0.085. Therefore, the average redshift was determined by the average redshift of the galaxies in the redshift range of $(z<0.085)$ in the ``superclean'' Galaxy Zoo and in \citep{shamir2020patterns}. In both cases the average redshift of the galaxies was $\sim$0.05, and therefore 0.05 was used as the average redshift of the galaxies in that dataset. Since the redshift range in \citep{longo2011detection} was $(z<0.085)$, the average redshift of the galaxies within the same range can be assumed to provide an approximation of the average redshift of the galaxies in that dataset.

While all the datasets mentioned above were taken from SDSS, a comparison should be made to more than a single telescope to show that if such asymmetry exists, it is not the feature of a single instrument or photometric pipeline. Relevant sky surveys that are not SDSS but collect large data and covering a sufficiently large footprint can be Pan-STARRS \citep{shamir2020patterns} and DESI Legacy Survey \citep{shamir2021large}. As shown in previous work, the location of the most likely dipole axis changes consistently when the redshift of the galaxies gets higher \citep{shamir2020patterns,shamir2022new}. That is, datasets that have similar redshift distribution tend to provide similar profiles of asymmetry, while the location of the peak of the axis changes when the redshift of the galaxies change \citep{shamir2020patterns,shamir2022new}. That can be viewed as an indication that if such axis exists, it does not necessarily go directly through Earth, as discussed in \citep{shamir2022new} and also briefly in Section~\ref{conclusion}. 



Table~\ref{comparison} summarizes the most likely peak shown in Section~\ref{reanalysis} for the dataset used in \citep{iye2020spin}, and the peak if the axis detected in other datasets mentioned above. While the locations of the peaks are not identical, they are within 1$\sigma$ error from the peak of the axis analyzed in Section~\ref{reanalysis}. The table also shows the number of galaxies in each dataset. The Pan-STARRS dataset is the smallest dataset, with merely $\sim3.3\cdot10^4$ galaxies, and the small size of the dataset can explain the lower statistical signal, slightly below 2$\sigma$. The largest dataset is the galaxies of the DESI Legacy Survey imaged by DECam, with over $8\cdot10^5$ galaxies.  The table also shows the number of galaxies annotated as clockwise galaxies and the number of galaxies annotated as counterclockwise galaxies in each dataset. As discussed in Section~\ref{error}, the sign of the asymmetry flips in opposite hemispheres, and therefore the asymmetry in one hemisphere is offset by the asymmetry in the opposite hemisphere. The total number of galaxies in the dataset that spin in opposite directions is therefore not an informative measurement of its asymmetry, as these numbers are heavily dependent on the dataset footprint. In the dataset of \cite{longo2011detection}, the number of galaxies spinning in each direction was determined from Figure 2 in \cite{longo2011detection}.



\begin{table*}[h]
\scriptsize
\begin{tabular}{|l|c|c|c|c|c|}
\hline
Telescope & Reference & RA          &    Dec         & Significance   &  \# galaxies \\ 
               &                  & (degrees) & (degrees)  &   ($\sigma$)  &  (cw | ccw)  \\
\hline
SDSS & This paper                                                             & 165 & 40 &  2.1   & 72,888 (36,607 | 36,181) \\
SDSS & \citep{land2008galaxy}                                           & 161 & 11 & $<2\sigma$ & 13,796 ( 6902 | 6,894) \\
SDSS & \citep{longo2011detection}                                     & 217   & 32 & 5.15 & 15,258 ( 7442 | 7,816 ) \\
DECam & \citep{shamir2021large}                                      & 237   & 10 & 4.7 & 807,898 (404,426 | 403,472) \\  
Pan-STARRS & \citep{shamir2020patterns}                         & 227   & 1   & 1.9 & 33,028 ( 16,508 | 16,520 ) \\
\hline
\end{tabular}
\caption{Most likely dipole axes from previous results using several different telescopes.}
\label{comparison}
\end{table*}

\section{Previous work that showed different conclusions}
\label{previous_studies}



Section~\ref{introduction} mentions briefly several studies that agree with the contention that the distribution of the spin directions of spiral galaxies is not necessarily random. That section also mentions some previous studies that proposed opposite conclusions. The purpose of this section is to analyze these studies and identify possible reasons for these conclusions.

One of the first studies to specifically claim that the spin directions of spiral galaxies is randomly distributed was done by \cite{iye1991catalog}. The dataset used to make the conclusions was a catalog of 6,525 galaxies (3,257 clockwise and 3,268 counterclockwise) from the Southern hemisphere. To observe a two-tailed statistical significance of P$\simeq$0.05, the distribution needs to be 3184:3341 or stronger. That is $\sim$5\% difference between the number of galaxies that spin clockwise and galaxies that spin counterclockwise. The asymmetry reported here is far smaller, at $\sim$1.4\%. That asymmetry is comparable or greater than the asymmetry reported in previous studies \citep{shamir2020patterns,shamir2021large}. Therefore, just $6.5\cdot10^3$ is not sufficient to identify the magnitude of asymmetry reported here or in previous studies. Based on binomial distribution when assuming the probability of a galaxy to spin in a certain direction is 0.5, showing a one-tailed statistically significant difference of P=0.05 when the difference between the number of clockwise and counterclockwise galaxies is 1.4\% requires a minimum population of 55,818 galaxies. Separation of such dataset to 27,715:28,103 would provide $\sim$1.4\% difference, and P$\simeq$0.05. Therefore, just a few thousand galaxies might not be sufficient to show statistically significant asymmetry.

Another study that aimed at addressing the same question and concluded that the distribution is random was done by using anonymous volunteers who annotated galaxy images by using a web-based user interface \citep{land2008galaxy}. While the annotation of a single anonymous untrained volunteer is not necessarily considered a reliable piece of information, a group of volunteers who annotate the same galaxy can provide a large number of annotations, and analysis of these annotations can provide meaningful information regrading the spin direction of the galaxy \citep{land2008galaxy}. One of the downsides of the study was that even a group of annotators does not necessarily guarantee accurate annotation. While each galaxy was annotated by multiple volunteers, the different annotations of the same galaxy do not always agree, making it difficult to know which annotations are the correct ones . That requires to reduce the data by defining a certain threshold majority as a criterion for the correctness of the annotation \citep{lintott2010}. For instance, for the separation between spiral and elliptical galaxies in Galaxy Zoo 1, just $\sim$40\% of the galaxies had agreement of at least 80\% of the annotators. That reduction led to a smaller dataset, but with a relatively small error of $\sim$3\% \citep{lintott2010}. To reduce the error rate to virtually 0\%, a ``superclean'' criterion of 95\% agreement was used, but just $\sim$12\% of the galaxies met that criterion \citep{lintott2010}. More importantly, the study also showed that the human perception is systematically biased \citep{land2008galaxy}.

The initial ``superclean'' dataset used in \citep{land2008galaxy} contained 6,106 galaxies spinning clockwise, and 7,034 galaxies spinning counterclockwise. That difference of $\sim$15\% was attributed to the perceptional bias of the human annotators \citep{land2008galaxy}. That bias was not known at the time of the design of the study, and therefore the galaxy images were initially not mirrored randomly to correct for that bias. When the bias was noticed, an experiment with mirrored galaxies was done for the purpose of profiling the bias, and included just a small subset of the data. That experiment showed that 5.525\% of the galaxies were classified as spinning clockwise, and 6.032\% of the galaxies were annotated as spinning counterclockwise. When the galaxies were mirrored, 5.942\% were annotated as clockwise, and 5.646\% as counterclockwise. While the experiment provided clear evidence of human annotation bias, it also showed that the results when mirroring the images were not identical to the results when using the original images. That is shown in Table 2 in \citep{land2008galaxy}. That shows that after mirroring the galaxies the number of counterclockwise galaxies was reduced by $\sim$1.5\%, and the number of clockwise galaxies increased by $\sim$2\%. That asymmetry is similar in direction and magnitude to the asymmetry shown in \citep{shamir2020patterns}. The observation reported in \citep{shamir2020patterns} is the most suitable comparison since it also analyzes SDSS galaxies with spectra, and therefore the footprint and distribution of the galaxies is similar.

The size of the entire set of galaxies that were mirrored in \citep{land2008galaxy} is 91,303. Based on that information, the number of galaxies annotated as clockwise was 5,044, and the number of mirrored galaxies classified as clockwise was 5155. The one-tailed binomial distribution statistics to get such asymmetry or stronger by chance is (P$<$0.13). While that P value is not considered statistically significant, there is certainty of $\sim$87\% for agreement with other reports on a higher number of counterclockwise galaxies among SDSS galaxies with spectra. When observing the galaxies classified as counterclockwise, the distribution is 5507:5425. That provides a P value of 0.21 to get such asymmetry or stronger by chance. Because the direction of asymmetry is in agreement in both clockwise and counterclockwise galaxies, the aggregated P value to have both results by chance is $\sim$0.03. That, however, is an invalid statistical analysis since the two experiments are not independent. While the annotations in each experiment are independent, it is likely that many of the galaxies that were annotated in both experiments were in fact the mirrored versions of the same galaxies.

As mentioned above, the asymmetry observed in \citep{land2008galaxy} agrees in both direction and magnitude with the asymmetry reported in \citep{shamir2020patterns} using SDSS galaxies with spectra, which means similar distribution of the objects within the SDSS footprint. The difference between the two experiments is the much larger size of the data used in \citep{shamir2020patterns}, which naturally provides a stronger statistical signal. It is therefore possible that a larger dataset of objects in \citep{land2008galaxy} would have shown a statistically significant signal. That is obviously an assumption that cannot be verified with the existing dataset of \citep{land2008galaxy}. But in any case, the two studies show results that are in agreement, and do not conflict with each other.

Another study that suggested that the distribution of spin directions of galaxies is random was done by using computer analysis to annotate the spin directions of the galaxies \citep{hayes2017nature}. The initial dataset was the same initial dataset of Galaxy Zoo, but instead of annotating the data manually, the researchers used the {\it SPARCFIRE} algorithm and software to identify the spin direction of the galaxies. The clear advantage of using a computer algorithm is that it is not subjected to the human perceptual bias that was dominant in the annotations performed by the Galaxy Zoo volunteers. {\it SPARCFIRE} is also a model-driven algorithm that is not based on machine learning, and therefore suitable for the task. Perhaps a possible limitation of the approach was that the process started with a first step of separating the spiral galaxies from the elliptical galaxies, which was done by using Galaxy Zoo human annotations of galaxies that were not mirrored randomly, or by machine learning. 

When applying the computer analysis to galaxies annotated as spirals by humans, the results showed asymmetry between the number of clockwise and counterclockwise galaxies with statistical significance as high as 2.84$\sigma$. That is shown in Table 2 in \citep{hayes2017nature}. That is, when using the galaxy images without making any prior assumptions about their distribution, the asymmetry in the distribution of clockwise and counterclockwise galaxies was statistically significant. 
That statistically significant observation was defined by the authors in the end of Section 3 of \citep{hayes2017nature} as ``surprising''.

A possible explanation would be that the humans who annotated the data were biased towards counterclockwise galaxies when they selected galaxies as spirals. That means that a spiral galaxy that spins counterclockwise is more likely to be selected as spiral compared to a spiral galaxy that spins clockwise. That is a new type of bias that was not known before. To correct for that possible bias, a machine learning algorithm was designed to select the galaxies in a manner that does not depend on its spin direction. The algorithm selected spiral galaxies from a much larger set of galaxies that contained both elliptical and spiral galaxies. The galaxies were selected by training a machine learning algorithm to differentiate between spiral and elliptical galaxies based on morphological descriptors computed by SPARCFIRE \citep{hayes2014}, as well as photometric attributes that allow separation between elliptical and spiral galaxies \citep{banerji2010}. To make the machine learning algorithm symmetric in its selection of spiral galaxies, the machine learning algorithm was trained after rejecting all attributes that could identify between clockwise and counterclockwise galaxies reported in \citep{shamir2016asymmetry}, as well as several other attributes used by {\it SPARCFIRE} that were noticed to differentiate between galaxies with opposite spin directions. 

 included in the dataset, which will discriminate galaxies with informative features that are different between clockwise and counterclockwise galaxies. Regardless of whether the source of the asymmetry is the real sky or a certain bias, the higher priority for attributes that are known as attributes that are not linked to the spin direction will reduce the asymmetry in the predictions compared to the original dataset. 


The random forest algorithm has an integrated feature selection process \citep{breiman2001}, and therefore it selects the informative features even when no explicit feature selection algorithm is used. A simple experiment was done by creating a random forest classifier that can differentiate between spiral and elliptical galaxies. For that purpose, a dataset of 19,500 spiral galaxies were taken from the dataset described in Section~\ref{reanalysis}, such that half of them were clockwise galaxies and the other half were galaxies spinning counterclockwise. That set of galaxies made the training samples of the spiral galaxies. The training samples of the elliptical galaxies were taken from the catalog of \citep{kuminski2016computer}, and included 19,500 annotated as elliptical with certainty of 0.9 or higher. Using the full photometric attributes of each galaxy taken from SDSS DR 8, a classifier $C_1$ was trained using random forest with the default settings of the {\it Scikit-learn} library.

The classifier $C_1$ was then applied to annotate a dataset of 10,000 clockwise galaxies and 9,500 counterclockwise galaxies used in Section~\ref{reanalysis}. After applying classifier $C_1$ to these galaxies, 17,169 galaxies were classified as spiral, and the rest of the galaxies were classified by the algorithm incorrectly as elliptical. The galaxies classified as spiral were distributed such that 8,814 galaxies had clockwise spin patterns, and 8,355 galaxies had counterclockwise spin patterns. That shows that among the galaxies classified by $C_1$ as spiral, there were $\sim$5.2\% less galaxies that spin counterclockwise. That asymmetry is very close to the asymmetry in the original dataset of 10,000 clockwise galaxies and 9,500 counterclockwise galaxies, which has exactly 5\% less galaxies that spin counterclockwise compared to the number of galaxies that spin clockwise. The classifier $C_1$ therefore did not lead to annotations that changed the distribution of the spin directions of the galaxies in the dataset that it classified.

Then, a classifier $C_2$ was trained with the same training set that was used to train classifier $C_1$. But for training classifier $C_2$, the 18 attributes from Table 8 in \citep{shamir2016asymmetry} were removed, in addition to isoPhiGrad attributes in all bands, isoPhi, petroR50Err attributes in all bands, petroR90Err attributes in all bands, u and q attributes in all bands, LnDeV attributes in all bands, lnLStar in all bands, and the magnitude attributes. After training $C_2$ with the same random forest using {\it Scikit-learn}, the classifier $C_2$ was applied to annotate the same 10,000 clockwise galaxies and 9,500 counterclockwise galaxies that were annotated by classifier $C_1$. The $C_2$ classifier annotated 14,541 galaxies as spiral, and the rest of the galaxies were incorrectly annotated by $C_2$ as elliptical. The set of galaxies that were classified as spiral by $C_2$ included 7,378 clockwise galaxies and 7,163 counterclockwise galaxies. Additionally, the resulting dataset of galaxies classified as spirals contained 3,814 elliptical galaxies that were incorrectly classified by the algorithm as spiral. While the resulting dataset still had a statistically significant higher number of galaxies spinning clockwise, the difference became smaller compared to the original dataset, from 5\% to $\sim$3\%. The two-tailed P-value of the binomial distribution of the asymmetry in the annotations of classifier $C_1$ (8,814 clockwise, 8,355 counterclockwise) is 0.00047. That probability increases to 0.076 with the annotations done by classifier $C_2$ (7,378 clockwise, 7,163 counterclockwise). That shows that when removing the attributes that differentiate between clockwise and counterclockwise galaxies, a dataset that was originally asymmetric became less asymmetric after applying a classifier to identify spiral galaxies. 

Because the machine learning algorithm used by \cite{hayes2017nature} was trained with an equal number of clockwise and counterclockwise spiral galaxies, there is no obvious reason for a machine learning algorithm to prefer one spin direction over the other even without removing certain attributes that correlate with the spin direction. The balanced training set is expected to ensure that the selection of spiral galaxies is not biased towards a certain spin direction.  But since machine learning is often not fully explainable, it is difficult to prove mathematically whether a certain machine learning algorithm is biased. That is often done by empirical analysis, which can also be challenging and non-intuitive as shown in \citep{dhar2022systematic}. Selecting the spiral galaxies using a machine learning algorithm provided a symmetric dataset after applying a removal of some of the attributes that correlate with the spin direction. As shown in the experiment above, the removal of such attributes can make a dataset with asymmetric distribution of galaxy spin directions become less asymmetric, and the asymmetry becomes statistically insignificant. While this experiment cannot be considered a proof that the removal of attributes led to the disappearance of the asymmetry in \citep{hayes2017nature}, the experiment above shows that it can reduce it, and therefore a possible explanation.

\cite{hayes2017nature} do not show the results after selecting the galaxies with their machine learning algorithm, yet without removing attributes. It is therefore not impossible that the asymmetry shown in this paper was observed also by \cite{hayes2017nature}, but was not included in their paper. Because the selection of spiral galaxies was done by training a machine learning system in a symmetric manner, the analysis makes a correct use of machine learning, within the known limitations of machine learning systems.

It should be mentioned that the attributes identified in \citep{shamir2016asymmetry} were reported as attributes that exhibit certain differences between clockwise and counterclockwise galaxies in SDSS. These attributes were never used for any analysis of the distribution of galaxies with opposite spin directions. The only known use of these attributes for that purpose is \citep{hayes2017nature}. As described in Section~\ref{selection_bias}, previous experiments were done by selecting all objects subjected to simple criteria (e.g., maximum radius), without making prior assumptions for the selection. By avoiding a first step of separation of the galaxies to elliptical and spiral, machine learning is not used, and its potential biases therefore cannot impact any stage of the analysis.




\section{Conclusion}
\label{conclusion}

As claimed in \citep{iye2020spin}, the distribution of spin directions of spiral galaxies is still unknown. Multiple experiments using several different telescopes suggest that the distribution might not be random. These include SDSS \citep{shamir2020large}, Pan-STARRS \citep{shamir2020patterns}, HST \citep{shamir2020pasa}, and DESI Legacy Survey \citep{shamir2021large,shamir2022new}. \cite{iye2020spin} used a dataset of photometric objects \citep{shamir2017photometric} that was used for an experiment aiming at profiling photometric differences between objects spinning in opposite directions, and used the same dataset to identify a cosmological-scale dipole in the S/Z galaxy distribution. They found that after removing photometric objects that are part of the same galaxy the dataset provided random distribution of 0.29$\sigma$ in the galaxy spin directions.

However, the random distribution of 0.29$\sigma$ was reported by limiting the dataset to $Z_{phot}<0.1$. The random distribution of galaxies in that redshift range agrees with previous literature showing random distribution of the spin directions in that redshift range \citep{shamir2020patterns}. More importantly, 
the analysis of \cite{iye2020spin} is three-dimensional, which requires the redshift of the galaxies. The source of the redshift is \citep{paul2018catalog}, which is a catalog of photometric redshift that \cite{iye2020spin}. 

Analysis of the exact same catalog used by \citep{iye2020spin}, but without limiting to low redshifts and without using the photometric redshift shows different statistical signal compared to the signal reported in \citep{iye2020spin}. That shows that by avoiding the use of the photometric redshift, the statistical signal of the same dataset is different. Because the analysis done in \citep{iye2020spin} is completely different from the analysis used here or in previous work \citep{shamir2012handedness,shamir2020patterns,shamir2021large}, it is not necessarily certain that limiting the redshift of the galaxies or using the photometric redshift are the reasons for the low statistical signal observed by \citep{iye2020spin}. Given that standard statistical methodology such as binomial distribution and $\chi^2$ show that the spin directions are not distributed randomly, the redshift limit and the use of photometric redshift are both possible reasons for the low statistical signal observed by \cite{iye2020spin}. Simple analyses show that limiting the data by the photometric redshift, or adding error to the position of each galaxy lead to the loss of the statistical signal of the dipole axis, even if the error is added in a symmetric manner.

The analysis shown here shows a peak that agrees with several previous experiments by several different researchers. The statistical significance, although greater that 2$\sigma$, is still not exceptionally high, and cannot prove or disprove the existence of a dipole axis. That is, the dataset used here does not necessarily prove that the distribution of the spin directions of SDSS galaxies forms a dipole axis. But it also does not show that the distribution of spin directions in SDSS is random. These results are added to previous work with other datasets, or with new analysis of datasets published in the past. That leads to the possibility that spin directions of objects in SDSS identified as galaxies are not necessarily random.


Previous work showed that the asymmetry in the spin directions of spiral galaxies can be observed in different telescopes, and the locations of the dipole axes observed with different telescopes are well within statistical error from each other. In addition to SDSS, the asymmetry was also observed in HST \citep{shamir2020pasa}, Pan-STARRS \citep{shamir2020patterns,shamir2021large,shamir2022new}, and DESI Legacy Survey \citep{shamir2021large,shamir2022new}.

Another interesting observation is that the location of the most likely axis depends on the redshift of the galaxies in the dataset \citep{shamir2020patterns,shamir2020pasa,shamir2022new}. That can be viewed as an indication that if such axis indeed exists, it might not necessarily go directly through Earth. When using datasets with similar redshift distribution of the galaxies, the location of the dipole axes observed with the different telescopes become closer, and the profile of the distribution of the galaxy spin directions becomes more similar \citep{shamir2020patterns,shamir2020pasa,shamir2022new}.

\begin{figure}[h!]
\includegraphics[scale=0.24]{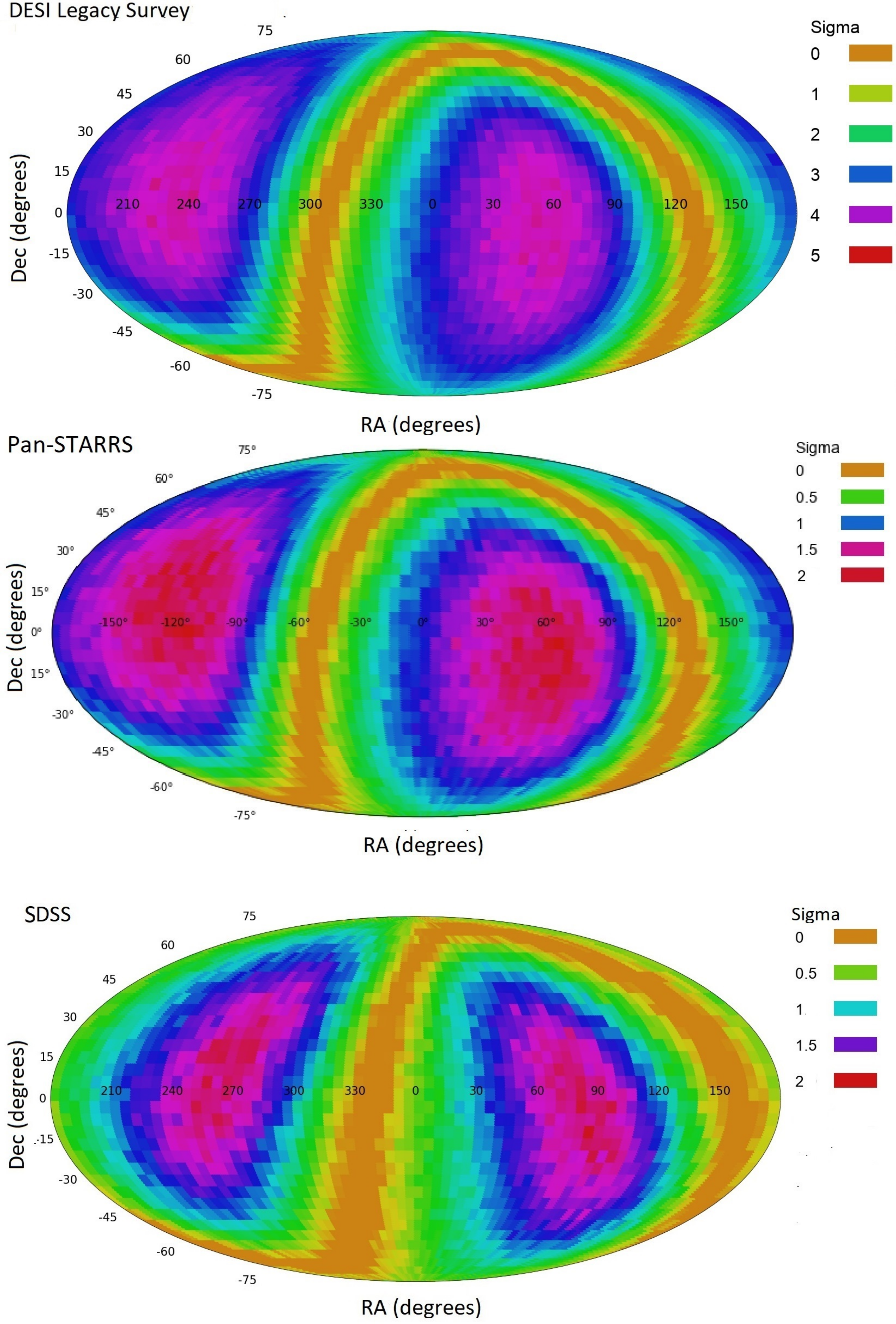}
\caption{The $\chi^2$ probability of the dipole axis in different $(\alpha,\delta)$ combinations when using galaxies in SDSS \citep{shamir2022new}, Pan-STARRS  \citep{shamir2020patterns}, and the DESI Legacy Survey \citep{shamir2021large}.}
\label{decam_sdss_panstarrs_normalized}
\end{figure}

Figure~\ref{decam_sdss_panstarrs_normalized} shows the results of the analysis described in Section~\ref{reanalysis} when using data from SDSS, Pan-STARRS \citep{shamir2020patterns}, and a relatively large dataset of $8.08\cdot10^5$ galaxies from the DESI Legacy Survey \citep{shamir2021large}. As described in \citep{shamir2022new}, the SDSS dataset was compiled by selecting $3.8\cdot10^4$ galaxies from the dataset used in \citep{shamir2020patterns} such that their redshift distribution is similar to the redshift distribution of the galaxies in the DESI Legacy Survey. Because most galaxies in DESI Legacy Survey do not yet have spectra, the redshift distribution of these galaxies was determined by a subset of $1.7\cdot10^4$ galaxies that had spectra through the 2dF survey \citep{cole20052df}. Full description of the normalization of the redshift distribution and the full-sky analysis can be found in \citep{shamir2022new}.

The figure shows very similar profiles of asymmetry across the different digital sky surveys, with similar locations of the most likely dipole axes \citep{shamir2022new}. The Pan-STARRS dataset of $\sim3.3\cdot10^4$ provided a dipole axis with statistical signal of 1.9$\sigma$, which might not necessarily be considered statistically significant, but also does not conflict with the other datasets. The SDSS galaxies $3.8\cdot10^4$ provided a statistical significance of 2.2$\sigma$ for the existence of the dipole axis, while the far larger dataset from DESI Legacy Survey of $8.08\cdot10^5$ galaxies showed a dipole axis with probability of 4.7$\sigma$ \citep{shamir2022new}.

When not normalizing the redshift distribution of the SDSS galaxies the most probable location of the dipole axis is still within 1$\sigma$ statistical error from the axes identified in Pan-STARRS and DESI Legacy Survey \citep{shamir2022new}. As described in \citep{shamir2020patterns,shamir2021large,shamir2022new}, assigning the spin directions with random distribution provided no statistically significant dipole axis, and a distribution profile similar to Figure~\ref{dipole_random}.
Interestingly, the locations of the most likely axes are very close to the CMB Cold Spot. For instance, the most likely axis observed in the DESI Legacy Survey data peaks at $(\alpha=57^o,\delta=-10^o)$, which is close to the CMB Cold Spot at $(\alpha=49^o,\delta=-19^o)$. That can obviously be considered a coincidence. The consistent change of the location of the most likely axis with the redshift can be viewed as an axis that does not necessarily go directly through Earth \citep{shamir2022new}.

The contention that the Universe is oriented around a major axis shifts from the standard cosmological models, but agrees with several other previously proposed cosmological theories. These include rotating universe \citep{godel1949example,ozsvath1962finite,ozsvath2001approaches,sivaram2012primordial,chechin2016rotation,seshavatharam2020integrated,camp2021}, ellipsoidal universe \citep{campanelli2006ellipsoidal,campanelli2007cosmic,campanelli2011cosmic,gruppuso2007complete,cea2014ellipsoidal}, or geometric inflation \citep{arciniega2020geometric,edelstein2020aspects,arciniega2020towards,jaime2021viability}. 

Another cosmological theory that could be relevant to the observation is Black Hole Cosmology \citep{pathria1972universe,stuckey1994observable,easson2001universe,poplawski2010radial,tatum2018clues,chakrabarty2020toy}, suggesting that the Universe is the interior of a black hole in another Universe. Black hole cosmology was motivated by the agreement between the Hubble radius and the Schwarzschild radius, but can also explain the accelerated expansion of the Universe without the assumption of dark energy. Black hole cosmology is also closely related to the theory of holographic universe \citep{susskind1995world,bak2000holographic,bousso2002holographic,myung2005holographic,hu2006interacting,sivaram2013holography,shor2021representation,rinaldi2022matrix}, which is related to black hole thermodynamics, and can explain space as seen in our Universe as the interior of a black hole in another universe. Because black holes spin \citep{gammie2004black,takahashi2004shapes,volonteri2005distribution,mcclintock2006spin,mudambi2020estimation,reynolds2021observational}, a universe hosted in a black hole should have an axis and a preferred direction inherited from the spin of its host black hole \citep{poplawski2010cosmology,seshavatharam2010physics,seshavatharam2014understanding,christillin2014machian,seshavatharam2020integrated}.

Large-scale anisotropy and a cosmological-scale axis were also proposed and discussed in the light of the cosmic microwave background radiation \citep{abramo2006anomalies,mariano2013cmb,land2005examination,ade2014planck,santos2015influence,dong2015inflation,gruppuso2018evens,yeung2022directional}, and the acceleration rates \citep{perivolaropoulos2014large}. \cite{luongo2022larger} proposed a link between the $H_0$ expansion rates and the CMB dipole. Other messengers that were used to show cosmological anisotropy in addition to the cosmic microwave background \citep{eriksen2004asymmetries,cline2003does,gordon2004low,campanelli2007cosmic,zhe2015quadrupole} include short gamma ray bursts \citep{meszaros2019oppositeness}, LX-T scaling \citep{migkas2020probing}, Ia supernova \citep{javanmardi2015probing,lin2016significance}, radio sources \citep{ghosh2016probing,tiwari2015dipole}, galaxy morphology types \citep{javanmardi2017anisotropy}, dark energy \citep{adhav2011kantowski,adhav2011lrs,perivolaropoulos2014large,colin2019evidence}, high-energy cosmic rays \citep{aab2017observation}, polarization of quasars \citep{hutsemekers2005mapping,secrest2021test}, and very large cosmological-scale structures \citep{deng2006super}. These observations challenge the standard cosmological models, and mandate further research to fully profile and understand their nature in the light of the large-scale structure of the Universe.


\section*{Acknowledgments}

I would like to thank the very knowledgeable anonymous reviewer for the sincere efforts to help improve the manuscript and the research. This study was supported in part by NSF grants AST-1903823 and IIS-1546079. 


\bibliographystyle{apalike}
\bibliography{main_archive}

\begin{thebibliography}{}

\bibitem[Aab et~al., 2017]{aab2017observation}
Aab, A., Abreu, P., Aglietta, M., Al~Samarai, I., Albuquerque, I., Allekotte,
  I., Almela, A., Castillo, J.~A., Alvarez-Mu{\~n}iz, J., Anastasi, G.~A.,
  et~al. (2017).
\newblock Observation of a large-scale anisotropy in the arrival directions of
  cosmic rays above 8$\times$ 1018 ev.
\newblock {\em Science}, 357(6357):1266--1270.

\bibitem[Abramo et~al., 2006]{abramo2006anomalies}
Abramo, L.~R., Sodr{\'e}~Jr, L., and Wuensche, C.~A. (2006).
\newblock Anomalies in the low cmb multipoles and extended foregrounds.
\newblock {\em Physical Review D}, 74(8):083515.

\bibitem[Ade et~al., 2014]{ade2014planck}
Ade, P.~A., Aghanim, N., Armitage-Caplan, C., Arnaud, M., Ashdown, M.,
  Atrio-Barandela, F., Aumont, J., Baccigalupi, C., Banday, A.~J., Barreiro,
  R., et~al. (2014).
\newblock Planck 2013 results. xxiii. isotropy and statistics of the cmb.
\newblock {\em Astronomy and Astrophysics}, 571:A23.

\bibitem[Adhav, 2011]{adhav2011lrs}
Adhav, K. (2011).
\newblock Lrs bianchi type-i universe with anisotropic dark energy in lyra
  geometry.
\newblock {\em International Journal of Astronomy and Astrophysics},
  1(4):204--209.

\bibitem[Adhav et~al., 2011]{adhav2011kantowski}
Adhav, K., Bansod, A., Wankhade, R., and Ajmire, H. (2011).
\newblock Kantowski-sachs cosmological models with anisotropic dark energy.
\newblock {\em Open Physics}, 9(4):919--925.

\bibitem[Arciniega et~al., 2020a]{arciniega2020geometric}
Arciniega, G., Bueno, P., Cano, P.~A., Edelstein, J.~D., Hennigar, R.~A., and
  Jaime, L.~G. (2020a).
\newblock Geometric inflation.
\newblock {\em Physics Letters B}, 802:135242.

\bibitem[Arciniega et~al., 2020b]{arciniega2020towards}
Arciniega, G., Edelstein, J.~D., and Jaime, L.~G. (2020b).
\newblock Towards geometric inflation: the cubic case.
\newblock {\em Physics Letters B}, 802:135272.

\bibitem[Bak and Rey, 2000]{bak2000holographic}
Bak, D. and Rey, S.-J. (2000).
\newblock Holographic principle and string cosmology.
\newblock {\em Classical and Quantum Gravity}, 17(1):L1.

\bibitem[Banerji et~al., 2010]{banerji2010}
Banerji, M., Lahav, O., Lintott, C.~J., Abdalla, F.~B., Schawinski, K.,
  Bamford, S.~P., Andreescu, D., Murray, P., Raddick, M.~J., Slosar, A.,
  Szalay, A., Thomas, D., and Vandenberg, J. (2010).
\newblock Galaxy zoo: reproducing galaxy morphologies via machine learning*.
\newblock {\em Monthly Notices of the Royal Astronomical Society},
  406(1):342--353.

\bibitem[Bernstein and Huterer, 2010]{bernstein2010catastrophic}
Bernstein, G. and Huterer, D. (2010).
\newblock Catastrophic photometric redshift errors: weak-lensing survey
  requirements.
\newblock {\em Monthly Notices of the Royal Astronomical Society},
  401(2):1399--1408.

\bibitem[Bousso, 2002]{bousso2002holographic}
Bousso, R. (2002).
\newblock The holographic principle.
\newblock {\em Reviews of Modern Physics}, 74(3):825.

\bibitem[Breiman, 2001]{breiman2001}
Breiman, L. (2001).
\newblock Random forests.
\newblock {\em Machine Learning}, 45(1):5--32.

\bibitem[Campanelli, 2021]{camp2021}
Campanelli, L. (2021).
\newblock A conjecture on the neutrality of matter.
\newblock {\em Foundations of Physics}, 51:56.

\bibitem[Campanelli et~al., 2011]{campanelli2011cosmic}
Campanelli, L., Cea, P., Fogli, G., and Tedesco, L. (2011).
\newblock Cosmic parallax in ellipsoidal universe.
\newblock {\em Modern Physics Letters A}, 26(16):1169--1181.

\bibitem[Campanelli et~al., 2006]{campanelli2006ellipsoidal}
Campanelli, L., Cea, P., and Tedesco, L. (2006).
\newblock Ellipsoidal universe can solve the cosmic microwave background
  quadrupole problem.
\newblock {\em Physical Review Letters}, 97(13):131302.

\bibitem[Campanelli et~al., 2007]{campanelli2007cosmic}
Campanelli, L., Cea, P., and Tedesco, L. (2007).
\newblock Cosmic microwave background quadrupole and ellipsoidal universe.
\newblock {\em Physical Review D}, 76(6):063007.

\bibitem[Cea, 2014]{cea2014ellipsoidal}
Cea, P. (2014).
\newblock The ellipsoidal universe in the planck satellite era.
\newblock {\em Monthly Notices of the Royal Astronomical Society},
  441(2):1646--1661.

\bibitem[Chakrabarty et~al., 2020]{chakrabarty2020toy}
Chakrabarty, H., Abdujabbarov, A., Malafarina, D., and Bambi, C. (2020).
\newblock A toy model for a baby universe inside a black hole.
\newblock {\em European Physical Journal C}, 80(1909.07129):1--10.

\bibitem[Chechin, 2016]{chechin2016rotation}
Chechin, L. (2016).
\newblock Rotation of the universe at different cosmological epochs.
\newblock {\em Astronomy Reports}, 60(6):535--541.

\bibitem[Christillin, 2014]{christillin2014machian}
Christillin, P. (2014).
\newblock The machian origin of linear inertial forces from our gravitationally
  radiating black hole universe.
\newblock {\em The European Physical Journal Plus}, 129(8):1--3.

\bibitem[Cline et~al., 2003]{cline2003does}
Cline, J.~M., Crotty, P., and Lesgourgues, J. (2003).
\newblock Does the small cmb quadrupole moment suggest new physics?
\newblock {\em Journal of Cosmology and Astroparticle Physics}, 2003(09):010.

\bibitem[Cole et~al., 2005]{cole20052df}
Cole, S., Percival, W.~J., Peacock, J.~A., Norberg, P., Baugh, C.~M., Frenk,
  C.~S., Baldry, I., Bland-Hawthorn, J., Bridges, T., Cannon, R., et~al.
  (2005).
\newblock The 2df galaxy redshift survey: power-spectrum analysis of the final
  data set and cosmological implications.
\newblock {\em Monthly Notices of the Royal Astronomical Society},
  362(2):505--534.

\bibitem[Colin et~al., 2019]{colin2019evidence}
Colin, J., Mohayaee, R., Rameez, M., and Sarkar, S. (2019).
\newblock Evidence for anisotropy of cosmic acceleration.
\newblock {\em Astronomy and Astrophysics}, 631:L13.

\bibitem[Dahlen et~al., 2013]{dahlen2013critical}
Dahlen, T., Mobasher, B., Faber, S.~M., Ferguson, H.~C., Barro, G.,
  Finkelstein, S.~L., Finlator, K., Fontana, A., Gruetzbauch, R., Johnson, S.,
  et~al. (2013).
\newblock A critical assessment of photometric redshift methods: a candels
  investigation.
\newblock {\em Astrophysical Journal}, 775(2):93.

\bibitem[Davis and Hayes, 2014]{hayes2014}
Davis, D.~R. and Hayes, W.~B. (2014).
\newblock {SpArcFiRe: Scalable Automated Detection of Spiral Galaxy Arm
  Segments}.
\newblock {\em Astrophysical Journal}, 790(2):87.

\bibitem[Deng et~al., 2006]{deng2006super}
Deng, X.-F., Chen, Y.-Q., Zhang, Q., and He, J.-Z. (2006).
\newblock Super-large-scale structures in the sloan digital sky survey.
\newblock {\em Chinese Journal of Astronomy and Astrophysics}, 6(1):35.

\bibitem[Dhar and Shamir, 2021]{dhar2021evaluation}
Dhar, S. and Shamir, L. (2021).
\newblock Evaluation of the benchmark datasets for testing the efficacy of deep
  convolutional neural networks.
\newblock {\em Visual Informatics}, 5(3):92--101.

\bibitem[Dhar and Shamir, 2022]{dhar2022systematic}
Dhar, S. and Shamir, L. (2022).
\newblock Systematic biases when using deep neural networks for annotating
  large catalogs of astronomical images.
\newblock {\em Astronomy and Computing}, 38:100545.

\bibitem[Dong et~al., 2015]{dong2015inflation}
Dong, Z., Ming-Hua, L., Ping, W., and Zhe, C. (2015).
\newblock Inflation in de sitter spacetime and cmb large scale anomaly.
\newblock {\em Chinese Physics C}, 39(9):095101.

\bibitem[Easson and Brandenberger, 2001]{easson2001universe}
Easson, D.~A. and Brandenberger, R.~H. (2001).
\newblock Universe generation from black hole interiors.
\newblock {\em Journal of High Energy Physics}, 2001(06):024.

\bibitem[Edelstein et~al., 2020]{edelstein2020aspects}
Edelstein, J.~D., Rodr{\'\i}guez, D.~V., and L{\'o}pez, A.~V. (2020).
\newblock Aspects of geometric inflation.
\newblock {\em Journal of Cosmology and Astroparticle Physics}, 2020(12):040.

\bibitem[Eriksen et~al., 2004]{eriksen2004asymmetries}
Eriksen, H.~K., Hansen, F.~K., Banday, A.~J., Gorski, K.~M., and Lilje, P.~B.
  (2004).
\newblock Asymmetries in the cosmic microwave background anisotropy field.
\newblock {\em Astrophysical Journal}, 605(1):14.

\bibitem[Gammie et~al., 2004]{gammie2004black}
Gammie, C.~F., Shapiro, S.~L., and McKinney, J.~C. (2004).
\newblock Black hole spin evolution.
\newblock {\em Astrophysical Journal}, 602(1):312.

\bibitem[Ghosh et~al., 2016]{ghosh2016probing}
Ghosh, S., Jain, P., Kashyap, G., Kothari, R., Nadkarni-Ghosh, S., and Tiwari,
  P. (2016).
\newblock Probing statistical isotropy of cosmological radio sources using
  square kilometre array.
\newblock {\em Journal of Astrophysics and Astronomy}, 37(4):1--21.

\bibitem[G{\"o}del, 1949]{godel1949example}
G{\"o}del, K. (1949).
\newblock An example of a new type of cosmological solutions of einstein's
  field equations of gravitation.
\newblock {\em Reviews of Modern Physics}, 21(3):447.

\bibitem[Gordon and Hu, 2004]{gordon2004low}
Gordon, C. and Hu, W. (2004).
\newblock Low cmb quadrupole from dark energy isocurvature perturbations.
\newblock {\em Physical Review D}, 70(8):083003.

\bibitem[Gruppuso, 2007]{gruppuso2007complete}
Gruppuso, A. (2007).
\newblock Complete statistical analysis for the quadrupole amplitude in an
  ellipsoidal universe.
\newblock {\em Physical Review D}, 76(8):083010.

\bibitem[Gruppuso et~al., 2018]{gruppuso2018evens}
Gruppuso, A., Kitazawa, N., Lattanzi, M., Mandolesi, N., Natoli, P., and
  Sagnotti, A. (2018).
\newblock The evens and odds of cmb anomalies.
\newblock {\em Physics of the Dark Universe}, 20:49--64.

\bibitem[Hayes et~al., 2017]{hayes2017nature}
Hayes, W.~B., Davis, D., and Silva, P. (2017).
\newblock On the nature and correction of the spurious s-wise spiral galaxy
  winding bias in galaxy zoo 1.
\newblock {\em Monthly Notices of the Royal Astronomical Society},
  466(4):3928--3936.

\bibitem[Hu and Ling, 2006]{hu2006interacting}
Hu, B. and Ling, Y. (2006).
\newblock Interacting dark energy, holographic principle, and coincidence
  problem.
\newblock {\em Physical Review D}, 73(12):123510.

\bibitem[Hutsem{\'e}kers et~al., 2005]{hutsemekers2005mapping}
Hutsem{\'e}kers, D., Cabanac, R., Lamy, H., and Sluse, D. (2005).
\newblock Mapping extreme-scale alignments of quasar polarization vectors.
\newblock {\em Astronomy and Astrophysics}, 441(3):915--930.

\bibitem[Iye and Sugai, 1991]{iye1991catalog}
Iye, M. and Sugai, H. (1991).
\newblock A catalog of spin orientation of southern galaxies.
\newblock {\em Astrophysical Journal}, 374:112--116.

\bibitem[Iye et~al., 2021]{iye2020spin}
Iye, M., Yagi, M., and Fukumoto, H. (2021).
\newblock Spin parity of spiral galaxies iii--dipole analysis of the
  distribution of sdss spirals with 3d random walk simulation.
\newblock {\em Astronomical Journal}, 907:123.

\bibitem[Jaime, 2021]{jaime2021viability}
Jaime, L.~G. (2021).
\newblock On the viability of the evolution of the universe with geometric
  inflation.
\newblock {\em Physics of the Dark Universe}, 34:100887.

\bibitem[Javanmardi and Kroupa, 2017]{javanmardi2017anisotropy}
Javanmardi, B. and Kroupa, P. (2017).
\newblock Anisotropy in the all-sky distribution of galaxy morphological types.
\newblock {\em Astronomy \& Astrophysics}, 597:A120.

\bibitem[Javanmardi et~al., 2015]{javanmardi2015probing}
Javanmardi, B., Porciani, C., Kroupa, P., and Pflam-Altenburg, J. (2015).
\newblock Probing the isotropy of cosmic acceleration traced by type ia
  supernovae.
\newblock {\em Astrophysical Journal}, 810(1):47.

\bibitem[Kuminski and Shamir, 2016]{kuminski2016computer}
Kuminski, E. and Shamir, L. (2016).
\newblock A computer-generated visual morphology catalog of ~3,000,000 sdss
  galaxies.
\newblock {\em Astrophysical Journal Supplement Series}, 223(2):20.

\bibitem[Land and Magueijo, 2005]{land2005examination}
Land, K. and Magueijo, J. (2005).
\newblock Examination of evidence for a preferred axis in the cosmic radiation
  anisotropy.
\newblock {\em Physical Review Letters}, 95(7):071301.

\bibitem[Land et~al., 2008]{land2008galaxy}
Land, K., Slosar, A., Lintott, C., Andreescu, D., Bamford, S., Murray, P.,
  Nichol, R., Raddick, M.~J., Schawinski, K., Szalay, A., et~al. (2008).
\newblock Galaxy zoo: the large-scale spin statistics of spiral galaxies in the
  sloan digital sky survey.
\newblock {\em Monthly Notices of the Royal Astronomical Society},
  388(4):1686--1692.

\bibitem[Lee et~al., 2019a]{lee2019galaxy}
Lee, J.~H., Pak, M., Lee, H.-R., and Song, H. (2019a).
\newblock Galaxy rotation coherent with the motions of neighbors: Discovery of
  observational evidence.
\newblock {\em Astrophysical Journal}, 872(1):78.

\bibitem[Lee et~al., 2019b]{lee2019mysterious}
Lee, J.~H., Pak, M., Song, H., Lee, H.-R., Kim, S., and Jeong, H. (2019b).
\newblock Mysterious coherence in several-megaparsec scales between galaxy
  rotation and neighbor motion.
\newblock {\em Astrophysical Journal}, 884(2):104.

\bibitem[Lin et~al., 2016]{lin2016significance}
Lin, H.-N., Li, X., and Chang, Z. (2016).
\newblock The significance of anisotropic signals hiding in the type ia
  supernovae.
\newblock {\em Monthly Notices of the Royal Astronomical Society},
  460(1):617--626.

\bibitem[Lintott et~al., 2010]{lintott2010}
Lintott, C., Schawinski, K., Bamford, S., Slosar, A., Land, K., Thomas, D.,
  Edmondson, E., Masters, K., Nichol, R.~C., Raddick, M.~J., Szalay, A.,
  Andreescu, D., Murray, P., and Vandenberg, J. (2010).
\newblock Galaxy zoo 1: data release of morphological classifications for
  nearly 900,000 galaxies.
\newblock {\em Monthly Notices of the Royal Astronomical Society},
  410(1):166--178.

\bibitem[Longo, 2011]{longo2011detection}
Longo, M.~J. (2011).
\newblock Detection of a dipole in the handedness of spiral galaxies with
  redshifts z~ 0.04.
\newblock {\em Physics Letters B}, 699(4):224--229.

\bibitem[Luongo et~al., 2022]{luongo2022larger}
Luongo, O., Muccino, M., Colg{\'a}in, E.~{\'O}., Sheikh-Jabbari, M., and Yin,
  L. (2022).
\newblock Larger h 0 values in the cmb dipole direction.
\newblock {\em Physical Review D}, 105(10):103510.

\bibitem[MacGillivray and Dodd, 1985]{macgillivray1985anisotropy}
MacGillivray, H. and Dodd, R. (1985).
\newblock The anisotropy of the spatial orientations of galaxies in the local
  supercluster.
\newblock {\em Astronomy and Astrophysics}, 145:269--274.

\bibitem[Mao et~al., 2017]{mao2017cosmic}
Mao, Q., Berlind, A.~A., Scherrer, R.~J., Neyrinck, M.~C., Scoccimarro, R.,
  Tinker, J.~L., McBride, C.~K., Schneider, D.~P., Pan, K., Bizyaev, D., et~al.
  (2017).
\newblock A cosmic void catalog of sdss dr12 boss galaxies.
\newblock {\em Astrophysical Journal}, 835(2):161.

\bibitem[Mariano and Perivolaropoulos, 2013]{mariano2013cmb}
Mariano, A. and Perivolaropoulos, L. (2013).
\newblock Cmb maximum temperature asymmetry axis: Alignment with other cosmic
  asymmetries.
\newblock {\em Physical Review D}, 87(4):043511.

\bibitem[McClintock et~al., 2006]{mcclintock2006spin}
McClintock, J.~E., Shafee, R., Narayan, R., Remillard, R.~A., Davis, S.~W., and
  Li, L.-X. (2006).
\newblock The spin of the near-extreme kerr black hole grs 1915+ 105.
\newblock {\em Astrophysical Journal}, 652(1):518.

\bibitem[M{\'e}sz{\'a}ros, 2019]{meszaros2019oppositeness}
M{\'e}sz{\'a}ros, A. (2019).
\newblock An oppositeness in the cosmology: Distribution of the gamma ray
  bursts and the cosmological principle.
\newblock {\em Astronomical Notes}, 340(7):564--569.

\bibitem[Migkas et~al., 2020]{migkas2020probing}
Migkas, K., Schellenberger, G., Reiprich, T., Pacaud, F., Ramos-Ceja, M., and
  Lovisari, L. (2020).
\newblock Probing cosmic isotropy with a new x-ray galaxy cluster sample
  through the lx--t scaling relation.
\newblock {\em Astronomy \& Astrophysics}, 636:A15.

\bibitem[Motloch et~al., 2021]{motloch2021observed}
Motloch, P., Yu, H.-R., Pen, U.-L., and Xie, Y. (2021).
\newblock An observed correlation between galaxy spins and initial conditions.
\newblock {\em Nature Astronomy}, 5(3):283--288.

\bibitem[Mudambi et~al., 2020]{mudambi2020estimation}
Mudambi, S.~P., Rao, A., Gudennavar, S., Misra, R., and Bubbly, S. (2020).
\newblock Estimation of the black hole spin in lmc x-1 using astrosat.
\newblock {\em Monthly Notices of the Royal Astronomical Society},
  498(3):4404--4410.

\bibitem[Myung, 2005]{myung2005holographic}
Myung, Y.~S. (2005).
\newblock Holographic principle and dark energy.
\newblock {\em Physics Letters B}, 610(1-2):18--22.

\bibitem[Orlov et~al., 2008]{orlov2008wnd}
Orlov, N., Shamir, L., Macura, T., Johnston, J., Eckley, D.~M., and Goldberg,
  I.~G. (2008).
\newblock Wnd-charm: Multi-purpose image classification using compound image
  transforms.
\newblock {\em Pattern Recognition Letters}, 29(11):1684--1693.

\bibitem[Ozsv{\'a}th and Sch{\"u}cking, 1962]{ozsvath1962finite}
Ozsv{\'a}th, I. and Sch{\"u}cking, E. (1962).
\newblock Finite rotating universe.
\newblock {\em Nature}, 193(4821):1168--1169.

\bibitem[Ozsvath and Sch{\"u}cking, 2001]{ozsvath2001approaches}
Ozsvath, I. and Sch{\"u}cking, E. (2001).
\newblock Approaches to g{\"o}del's rotating universe.
\newblock {\em Classical and Quantum Gravity}, 18(12):2243.

\bibitem[Pathria, 1972]{pathria1972universe}
Pathria, R. (1972).
\newblock The universe as a black hole.
\newblock {\em Nature}, 240(5379):298--299.

\bibitem[Paul et~al., 2018]{paul2018catalog}
Paul, N., Virag, N., and Shamir, L. (2018).
\newblock A catalog of photometric redshift and the distribution of broad
  galaxy morphologies.
\newblock {\em Galaxies}, 6(2):64.

\bibitem[Perivolaropoulos, 2014]{perivolaropoulos2014large}
Perivolaropoulos, L. (2014).
\newblock Large scale cosmological anomalies and inhomogeneous dark energy.
\newblock {\em Galaxies}, 2(1):22--61.

\bibitem[Pop{\l}awski, 2010a]{poplawski2010cosmology}
Pop{\l}awski, N.~J. (2010a).
\newblock Cosmology with torsion: An alternative to cosmic inflation.
\newblock {\em Physics Letters B}, 694(3):181--185.

\bibitem[Pop{\l}awski, 2010b]{poplawski2010radial}
Pop{\l}awski, N.~J. (2010b).
\newblock Radial motion into an einstein--rosen bridge.
\newblock {\em Physics Letters B}, 687(2-3):110--113.

\bibitem[Rau et~al., 2015]{rau2015accurate}
Rau, M.~M., Seitz, S., Brimioulle, F., Frank, E., Friedrich, O., Gruen, D., and
  Hoyle, B. (2015).
\newblock Accurate photometric redshift probability density estimation--method
  comparison and application.
\newblock {\em Monthly Notices of the Royal Astronomical Society},
  452(4):3710--3725.

\bibitem[Reynolds, 2021]{reynolds2021observational}
Reynolds, C.~S. (2021).
\newblock Observational constraints on black hole spin.
\newblock {\em Annual Review of Astronomy and Astrophysics}, 59:117--154.

\bibitem[Rinaldi et~al., 2022]{rinaldi2022matrix}
Rinaldi, E., Han, X., Hassan, M., Feng, Y., Nori, F., McGuigan, M., and Hanada,
  M. (2022).
\newblock Matrix-model simulations using quantum computing, deep learning, and
  lattice monte carlo.
\newblock {\em PRX Quantum}, 3(1):010324.

\bibitem[Santos et~al., 2015]{santos2015influence}
Santos, L., Cabella, P., Villela, T., and Zhao, W. (2015).
\newblock Influence of planck foreground masks in the large angular scale
  quadrant cmb asymmetry.
\newblock {\em Astronomy and Astrophysics}, 584:A115.

\bibitem[Secrest et~al., 2021]{secrest2021test}
Secrest, N.~J., von Hausegger, S., Rameez, M., Mohayaee, R., Sarkar, S., and
  Colin, J. (2021).
\newblock A test of the cosmological principle with quasars.
\newblock {\em ApJL}, 908(2):L51.

\bibitem[Seshavatharam, 2010]{seshavatharam2010physics}
Seshavatharam, U. (2010).
\newblock Physics of rotating and expanding black hole universe.
\newblock {\em Progress in Physics}, 2:7--14.

\bibitem[Seshavatharam and Lakshminarayana,
  2014]{seshavatharam2014understanding}
Seshavatharam, U. and Lakshminarayana, S. (2014).
\newblock Understanding black hole cosmology and the cosmic halt.
\newblock {\em Journal of Advanced Research in Astrophysics and Astronomy},
  1(1):1--27.

\bibitem[Seshavatharam and Lakshminarayana, 2020]{seshavatharam2020integrated}
Seshavatharam, U. and Lakshminarayana, S. (2020).
\newblock An integrated model of a light speed rotating universe.
\newblock {\em International Astronomy and Astrophysics Research Journal},
  pages 74--82.

\bibitem[Shamir, 2011]{shamir2011ganalyzer}
Shamir, L. (2011).
\newblock Ganalyzer: A tool for automatic galaxy image analysis.
\newblock {\em Astrophysical Journal}, 736(2):141.

\bibitem[Shamir, 2012]{shamir2012handedness}
Shamir, L. (2012).
\newblock Handedness asymmetry of spiral galaxies with z< 0.3 shows cosmic
  parity violation and a dipole axis.
\newblock {\em Physics Letters B}, 715(1-3):25--29.

\bibitem[Shamir, 2016]{shamir2016asymmetry}
Shamir, L. (2016).
\newblock Asymmetry between galaxies with clockwise handedness and
  counterclockwise handedness.
\newblock {\em Astrophysical Journal}, 823(1):32.

\bibitem[Shamir, 2017a]{shamir2017large}
Shamir, L. (2017a).
\newblock Large-scale photometric asymmetry in galaxy spin patterns.
\newblock {\em Publications of the Astronomical Society of Australia}, 34:e44.

\bibitem[Shamir, 2017b]{shamir2017photometric}
Shamir, L. (2017b).
\newblock Photometric asymmetry between clockwise and counterclockwise spiral
  galaxies in sdss.
\newblock {\em Publications of the Astronomical Society of Australia}, 34:e011.

\bibitem[Shamir, 2020a]{shamir2020asymmetry}
Shamir, L. (2020a).
\newblock Asymmetry between galaxies with different spin patterns: A comparison
  between cosmos, sdss, and pan-starrs.
\newblock {\em Open Astronomy}, 29(1):15--27.

\bibitem[Shamir, 2020b]{shamir2020pasa}
Shamir, L. (2020b).
\newblock Galaxy spin direction distribution in hst and sdss show similar
  large-scale asymmetry.
\newblock {\em Publications of the Astronomical Society of Australia}, 37:e053.

\bibitem[Shamir, 2020c]{shamir2020large}
Shamir, L. (2020c).
\newblock Large-scale asymmetry between clockwise and counterclockwise galaxies
  revisited.
\newblock {\em Astronomical Notes}, 341(3):324.

\bibitem[Shamir, 2020d]{shamir2020patterns}
Shamir, L. (2020d).
\newblock Patterns of galaxy spin directions in sdss and pan-starrs show parity
  violation and multipoles.
\newblock {\em ApSS}, 365:136.

\bibitem[Shamir, 2021a]{shamir2021particles}
Shamir, L. (2021a).
\newblock Analysis of the alignment of non-random patterns of spin directions
  in populations of spiral galaxies.
\newblock {\em Particles}, 4(1):11--28.

\bibitem[Shamir, 2021b]{shamir2021large}
Shamir, L. (2021b).
\newblock Large-scale asymmetry in galaxy spin directions: evidence from the
  southern hemisphere.
\newblock {\em Publications of the Astronomical Society of Australia}, 38:e037.

\bibitem[Shamir, 2022a]{shamir2022analysis}
Shamir, L. (2022a).
\newblock Analysis of $\sim10^6$ spiral galaxies from four telescopes shows
  large-scale patterns of asymmetry in galaxy spin directions.
\newblock {\em Advances in Astronomy}, 2022:8462363.

\bibitem[Shamir, 2022b]{shamir2022large}
Shamir, L. (2022b).
\newblock Large-scale asymmetry in galaxy spin directions: Analysis of galaxies
  with spectra in des, sdss, and desi legacy survey.
\newblock {\em Astronomical Notes}, page e20220010.

\bibitem[Shamir, 2022c]{shamir2022new}
Shamir, L. (2022c).
\newblock New evidence and analysis of cosmological-scale asymmetry in galaxy
  spin directions.
\newblock {\em Journal of Astrophysics and Astronomy}, page In Press. ArXiv:
  2201.03757.

\bibitem[Shor et~al., 2021]{shor2021representation}
Shor, O., Benninger, F., and Khrennikov, A. (2021).
\newblock Representation of the universe as a dendrogramic hologram endowed
  with relational interpretation.
\newblock {\em Entropy}, 23(5):584.

\bibitem[Sivaram and Arun, 2012]{sivaram2012primordial}
Sivaram, C. and Arun, K. (2012).
\newblock Primordial rotation of the universe, hydrodynamics, vortices and
  angular momenta of celestial objects.
\newblock {\em Open Astronomy}, 5:7--11.

\bibitem[Sivaram and Arun, 2013]{sivaram2013holography}
Sivaram, C. and Arun, K. (2013).
\newblock Holography, dark energy and entropy of large cosmic structures.
\newblock {\em ApSS}, 348(1):217--219.

\bibitem[Stuckey, 1994]{stuckey1994observable}
Stuckey, W. (1994).
\newblock The observable universe inside a black hole.
\newblock {\em American Journal of Physics}, 62(9):788--795.

\bibitem[Susskind, 1995]{susskind1995world}
Susskind, L. (1995).
\newblock The world as a hologram.
\newblock {\em Journal of Mathematical Physics}, 36(11):6377--6396.

\bibitem[Takahashi, 2004]{takahashi2004shapes}
Takahashi, R. (2004).
\newblock Shapes and positions of black hole shadows in accretion disks and
  spin parameters of black holes.
\newblock {\em Astrophysical Journal}, 611(2):996.

\bibitem[Tanaka et~al., 2018]{tanaka2018photometric}
Tanaka, M., Coupon, J., Hsieh, B.-C., Mineo, S., Nishizawa, A.~J., Speagle, J.,
  Furusawa, H., Miyazaki, S., and Murayama, H. (2018).
\newblock Photometric redshifts for hyper suprime-cam subaru strategic program
  data release 1.
\newblock {\em Publications of the Astronomical Society of Japan}, 70(SP1):S9.

\bibitem[Tatum et~al., 2018]{tatum2018clues}
Tatum, E.~T., Seshavatharam, U., et~al. (2018).
\newblock Clues to the fundamental nature of gravity, dark energy and dark
  matter.
\newblock {\em Journal of Modern Physics}, 9(08):1469.

\bibitem[Timmis and Shamir, 2017]{timmis2017catalog}
Timmis, I. and Shamir, L. (2017).
\newblock A catalog of automatically detected ring galaxy candidates in
  panstarss.
\newblock {\em Astrophysical Journal Supplement Series}, 231(1):2.

\bibitem[Tiwari and Jain, 2015]{tiwari2015dipole}
Tiwari, P. and Jain, P. (2015).
\newblock Dipole anisotropy in integrated linearly polarized flux density in
  nvss data.
\newblock {\em Monthly Notices of the Royal Astronomical Society},
  447(3):2658--2670.

\bibitem[Volonteri et~al., 2005]{volonteri2005distribution}
Volonteri, M., Madau, P., Quataert, E., and Rees, M.~J. (2005).
\newblock The distribution and cosmic evolution of massive black hole spins.
\newblock {\em Astrophysical Journal}, 620(1):69.

\bibitem[Watanabe, 2021]{Fukumoto2021}
Watanabe, J. (2021).
\newblock Private communication.
\newblock {\em Private communication}.

\bibitem[Wittman, 2009]{wittman2009lies}
Wittman, D. (2009).
\newblock What lies beneath: Using p (z) to reduce systematic photometric
  redshift errors.
\newblock {\em Astrophysical Journal}, 700(2):L174.

\bibitem[Yeung and Chu, 2022]{yeung2022directional}
Yeung, S. and Chu, M.-C. (2022).
\newblock Directional variations of cosmological parameters from the planck cmb
  data.
\newblock {\em Physical Review D}, 105:083508.

\bibitem[Zhe et~al., 2015]{zhe2015quadrupole}
Zhe, C., Xin, L., and Sai, W. (2015).
\newblock Quadrupole-octopole alignment of cmb related to the primordial power
  spectrum with dipolar modulation in anisotropic spacetime.
\newblock {\em Chinese Physics C}, 39(5):055101.

\end{thebibliography}

\end{document}